\documentclass[12pt]{article}

\usepackage{amsmath,amssymb}
\usepackage{graphicx}
\usepackage{epstopdf}
\usepackage{epsfig}
\usepackage{array,multirow}
\usepackage{subfigure}

\begin{document}


\title{
\hfill{\normalsize\vbox{%
\hbox{\rm 
}
\hbox{\rm 
}
}}
\vspace*{0.5cm}\\
{ Anomaly-Mediation and Sequestering \\
from a Higher-Dimensional Viewpoint}
\vspace{0.1cm}
}

\author{
{ {Minho Son\thanks{e-mail: mhson@pha.jhu.edu}} and
  {Raman Sundrum\thanks{e-mail: sundrum@pha.jhu.edu}} }\\ [0.7cm]
{\small \it Department of Physics and Astronomy, Johns Hopkins University}\\
{\small \it 3400 North Charles St., Baltimore, MD 21218}\vspace{0.1cm}
}

\date{}
\maketitle

\begin{abstract}
We study a five-dimensional supergravity model with boundary-localized visible sector exhibiting anomaly-mediated supersymmetry breaking, in which the central requirements of sequestering and radius stabilization are achieved perturbatively. This makes it possible to understand these various mechanisms in a more integrated and transparent fashion, mostly from the higher-dimensional viewpoint. Local supersymmetry, in the presence of visible sector quantum effects, is  enforced by the formalism of the five-dimensional superconformal tensor calculus. The construction results in only mild warping, which allows a natural supersymmetry-breaking mediation mechanism of (finite) boundary-to-boundary gravity loops to co-dominate with anomaly-mediation, thereby solving the latter's tachyonic slepton problem. We make the non-trivial check that this can occur while dangerous loops of stabilizing fields remain highly suppressed. Our discussion is a well-controlled starting point for considering other  generalizations of anomaly-mediation, or for string theory realizations.
\end{abstract}

\newpage


\section{Introduction}
Anomaly-Mediated Supersymmetry Breaking (AMSB) is an elegant but rather subtle Supersymmetry (SUSY) breaking mediation mechanism operating within supergravity theories~\cite{Randall:1999ee,GiudiceLMR:1998}. It offers an attractive solution to the supersymmetric flavor and CP problems of weak scale supersymmetry, and realistic models are possible when applied to either non-minimal supersymmetric extensions of the Standard Model (SM) or when combined with other SUSY breaking effects \cite{hybrid,nonmin}. Part of the subtlety is the important role  played by the quantum conformal anomaly (running couplings)  within the ``visible'' or Standard Model sector, when propagating in a supergravity background. Partly the subtlety involves the precondition for the dominance of AMSB effects in the visible sector, known as {\it sequestering}.

For perturbative visible sectors, AMSB is also a perturbative effect. In principle, sequestering can also take a perturbative form, when the visible sector and hidden sector are localized on separated ``branes''  in a higher-dimensional spacetime~\cite{Randall:1999ee}. Therefore one might expect that realistic models of AMSB could be built where all relevant calculations could be worked out by means of  Feynman diagrams and classical solutions in a higher-dimensional locally supersymmetric setting. In such a setting the subtleties of AMSB and sequestering could be understood most cleanly.   Perhaps surprisingly, this has proven difficult.

One forbidding aspect has been the complexity of higher-dimensional supergravity with bulk matter and their consistent couplings to the quantum SM. As a result, the original discussions of sequestering worked at the level of the 4D effective field theory below the compactification scale, with key properties being deduced from matching to the higher-dimensional theory~\cite{4DAMSB:flat,4DAMSB:warped}. This makes the AMSB-sequestering connection somewhat opaque, but it was possible to proceed in this manner. Even here, non-perturbative physics has been invoked in the 4D effective theory in order to achieve extra-dimensional radius stabilization, an issue entangled with AMSB. One would like to achieve a more microscopic, higher-dimensional, and perturbative view of what is going on, for the sake of greater clarity, certainty, and as the basis for further developments.

In the present paper, we study a perturbative 5D model with AMSB and sequestering which allows one to see them operate directly from the 5D perspective, with minimal recourse to the 4D effective theory below compactification.  One can consider it as either an effective field theory of a possible string theory compactification (with perhaps even more extra dimensions, but where the fifth dimension is the largest), or just as a simplified model. Given the complexity of full string theory constructions, 5D effective theories allow one to zoom in onto just the essential features of AMSB and sequestering. We hope that the 5D effective description we give is a useful basis for exploring, in the simplest possible manner, new mechanisms that can function symbiotically with AMSB. The basic 5D model is not new and was proposed in Ref.~\cite{MaruO:2004}, with somewhat different emphasis. However, our discussion has important new elements, in particular the care we give to tracking local supersymmmetry in the presence of the quantum conformal anomaly, and to AMSB and sequestering. We give a more explicit, complete and fully nonlinear treatment at 5D level.

Let us review the central issues of AMSB and sequestering in order to understand the path we will follow in the paper. We begin with a simple and familiar non-supersymmetric warm-up, namely pure non-abelian Yang-Mills theory in flat 4D spacetime,
\begin{equation}
{\cal L} = -\frac{1}{4 g^2} (F_{\mu \nu}^a)^2.
\end{equation}
Classically, this theory is exactly conformally invariant, but this symmetry is broken quantum mechanically by the running of the coupling. Now let us couple this theory to 4D gravity classically,
\begin{equation}
{\cal L} = \sqrt{-g}\, \{  - \frac{1}{4 g^2} g^{\mu \nu} g^{\alpha \beta} 
F_{\mu \alpha}^a F_{\nu \beta}^a \}.
\end{equation}
The action now enjoys general coordinate invariance, which contains conformal invariance as a subgroup in the following sense. Under a general coordinate transformation the Minkowski metric transforms as
\begin{equation}
ds^2 = \eta_{\mu \nu} dx^{\mu} dx^{\nu} \rightarrow
g'_{\mu \nu}(x') dx^{' \mu} dx^{' \nu},
\end{equation}
in the usual way. But for a subgroup  of such coordinate transformations, the transformed metric takes the form
\begin{equation}
g'_{\mu \nu}(x') = f^2(x') \eta_{\mu \nu}.
\end{equation}
This subgroup of coordinate transformations defines the conformal transformations, which therefore must be a symmetry of the action when coupled to the metric. 

At the quantum level, the question is how conformal invariance can be both exact,  being a subgroup of coordinate invariance, and also anomalous as is familiar in non-abelian Yang-Mills theory. The answer is that the scalar (off-shell) mode of the metric,
\begin{equation}
g_{\mu \nu}(x) \equiv C^2(x) \eta_{\mu \nu} + {\rm the ~rest},
\end{equation}
transforms under conformal transformations, $C \rightarrow f(x) C$, and must couple precisely so as to compensate for any conformal breaking in the non-gravitational dynamics. Let us focus on just this mode. It is straightforward to see that classically $C$-dependence cancels out of the gravitationally coupled Yang-Mills action. But at the quantum level this is not so. For example in dimensional regularization, in $4 + \epsilon$ dimensions, $C$ clearly no longer cancels out, and multiplies the renormalization scale $\mu$-dependence introduced to balance dimensions. After renormalization, exact coordinate conformal invariance implies the Yang-Mills dynamics in the $C(x)$ background takes the form 
\begin{equation}
{\cal L} = -\frac{1}{4 g^2(\mu C(x))} (F_{\mu \nu}^a)^2.
\end{equation}
In this way, the breaking of conformal invariance in Yang-Mills theory due to $\mu$-dependence is dressed up by the couplings to $C$ so as to appears as a spontaneous breaking in Minkowski space,
\begin{equation}
\langle C \rangle = 1.
\end{equation}

Deviations of C about this Vacuum Expectation Value (VEV), $C(x) \equiv 1 + \delta C(x)$, correspond to real spacetime curvature. For small and slowly varying $\delta C(x)$, the coupling to the Yang-Mills sector is given by Taylor expanding the running gauge coupling,
\begin{eqnarray}
{\cal L} = [-\frac{1}{4 g^2(\mu)} + \frac{\beta(g(\mu))}{2 g^3(\mu)} \delta C]
 (F_{\mu \nu}^a)^2 .
\end{eqnarray}
This is the bosonic equivalent of anomaly-mediation. See Ref. \cite{seiberg,alwis}
for other discussions of AMSB and analogous phenomena.

When we pass to supergravity, the scalar mode $C$ must be promoted to an off-shell chiral multiplet which compensates for superconformal breaking, explicit or anomalous, of the matter sector. It is the only supergravity mode that can get a non-zero VEV in Minkowski spacetime (i.e. Poincare invariant VEV),
\begin{equation}
\langle C \rangle = 1 + {\cal F}_{C}\, \theta^2.
\end{equation}
It is convenient (but not essential) that there is a supergravity formalism, the Superconformal Tensor Calculus \cite{KakuTN,Ferrara:1978rk,Kugo:1982cu,KugoU:1983}, that effectively gives special status to the compensator $C$ mode, and to the exact superconformal invariance of its couplings. If we are only interested in the propagation of a renormalizable matter sector (generalizing our Yang-Mills example) in such a Poincare-invariant supergravity background, then we can write the couplings to $C$ using flat superspace notation. Local supersymmetry demands that the couplings are exactly superconformally invariant,
\begin{equation}
\label{superconfinvcoup}
{\cal L} = \int d^4 \theta\, Z(\mu \sqrt{CC^*}) \, Q^* e^V Q + \int d^2 \theta\, W(Q,C) +
\int d^2 \theta\, \tau_{1-loop}(C \mu) {\cal W}_{\alpha}^2 +  {\rm h.c.}.
\end{equation}
Here, the superpotential is exactly cubic in fields to maintain classical conformal invariance and we are working in the field normalization such that the superpotential is not renormalized and the gauge coupling $\tau_{1-loop} \equiv 1/g^2$ is renormalized at only one loop. In this language all the remaining running is incorporated into the wavefunction renormalization. (This allows us to use holomorphicity to constrain the appearance of $C$ in $\tau$.) The standard results of AMSB result by expanding this equation in superspace, taking into account ${\cal F}_{C}\, \theta^2$ (Each of the original Refs.~\cite{Randall:1999ee,GiudiceLMR:1998} contain some but not all of these results. For example, the full combined list is reviewed in Ref.~\cite{nonmin}). To the extent that the $\mu$ dependence is dominated by gauge dynamics, the soft terms induced are flavor-blind.

However, in supergravity supersymmetry breaking ${\cal F}_C \neq 0$ can only occur in Minkowski space if accompanied by supersymmetry breaking from the matter sector, say a hidden sector. Then cancellation of the effective cosmological constant typically relates these two by
\begin{equation}
{\cal F}_C \sim {\cal F}_{hid}/M_{Pl}.
\end{equation}
Then there can be additional sources of SUSY breaking felt by the visible sector $Q,V$  via even Planck-suppressed couplings between the visible and hidden sectors, such as
\begin{equation}
\sim \int d^4 \theta\, \frac{\Sigma^* \Sigma}{M_{Pl}^2} 
Q^* e^V Q\, ,
\end{equation}
where $\Sigma$ denotes some hidden chiral fields. While the AMSB-induced visible soft terms are SM-loop suppressed, since they 
typically depend on the conformal anomaly, the soft terms directly from the hidden sector are not, and furthermore they may be highly flavor-violating. If such terms are present, AMSB is subdominant and does not provide a solution to the supersymmetric flavor problem.

A mechanism to further suppress even the Planck-suppressed hidden-visible couplings is known as "sequestering". We follow the original proposal of having the hidden and visible sectors localized on different 4D boundaries in a 5D supergravity theory, so that no direct couplings are allowed by locality~\cite{Randall:1999ee}. One must also check that hidden-visible couplings are then not induced by integrating out massive bulk modes. The study of these sequestering issues and AMSB directly in the higher-dimensional context is facilitated by the development in recent years of a 5D Superconformal Tensor Calculus for 5D supergravity~\cite{Kugo56:2002,FujitaO:2001} (with important developments earlier in~\cite{Zucher:offshell,Zucher:2002} as well) and matter~\cite{Kugo56:2002,FujitaO:2001,KugoMatter:2002,FujitaKO:2001,Bergshoeff:2002} and their couplings to 4D boundary fields~\cite{Zucher:offshellbrane,FujitaKO:2001,KugoOrbifolding:2002}. This formalism incorporates 5D superconformal bulk compensators whose boundary restrictions then play the role of the 4D compensator reviewed above. This is more transparent than earlier AMSB work~\cite{4DAMSB:flat,4DAMSB:warped} in which the compensator is introduced only after arriving at the 4D effective field theory below compactification.

The issue of radius stabilization is intertwined with AMSB and sequestering. For example in the no-scale model which emerges from a 5D set-up without stabilization, the effective 4D compensator has ${\cal F}_C = 0$, with vanishing AMSB~\cite{4DAMSB:flat}. In other examples, failure to stabilize results in runaway radius moduli after hidden sector SUSY breaking~\cite{4DAMSB:flat}. However, the requisite stabilizing fields in the bulk also raise this question: given that these fields are needed to get AMSB in the visible sector boundary from SUSY breaking on the hidden sector boundary, how is it that they do not mediate any unsequestered SUSY breaking effects as well? Of course, any answer should involve the constraints of local supersymmetry, but the understanding has
been necessarily indirect in earlier  studies of AMSB from higher dimensions because stabilization was accomplished by non-perturbative gaugino condensation, which can only be captured within the 4D effective field theory, while matching to this theory from 5D must be performed perturbatively~\cite{4DAMSB:flat}. In the present paper we adopt a transparent classical mechanism of radius stabililzation, by means of massive bulk hypermultiplet fields with boundary superpotentials, as proposed earlier in Ref.~\cite{MaruO:2004} (See Refs.~\cite{ArkaniHamed:1999pv,Eto:2004yk} for related work, but in absence of supergravity). The stabilization is essentially a supersymmetric generalization of the Goldberger-Wise mechanism~\cite{GoldbergerW:1999}.

A final consideration in AMSB is that when applied to the minimal visible sector, the Minimal Supersymmetric Standard Model (MSSM), it results in unacceptable tachyonic slepton masses. When the compactification scale is not too much smaller than the Planck scale, there is an attractive flavor-blind and UV-finite gravity loop effect that can generate non-tachyonic visible scalar masses that compete with AMSB, and therefore can result in a viable spectrum \cite{GregoireRS:2005} .\footnote{
This still does not solve the $\mu$ problem of the MSSM though.} However, this is only true of mildly warped compactifications. In highly warped compactifications the gravity loop effects are far subdominant to AMSB. We will therefore study the mildly warped case here. The highly warped case is interesting because it is AdS/CFT dual~\cite{Maldacena:1997} to the purely 4D mechanism of ``conformal sequestering'', which avoids the complications of higher-dimension supergravity, but requires non-perturbatively strong couplings in the hidden sector~\cite{Luty:2001jh,Luty:2001zv}. The highly warped case is also interesting because a fully sequestered example of this type has been established within string theory~\cite{Kachru:2007s}. We hope that our study of a mildly warped 5D model will help in finding mildly warped string theory constructions in which flavor-blind gravity loops work in conjunction with AMSB.

The paper is organized as follows. Section 2 contains a lightning introduction to the multiplets and component fields of 5D superconformal tensor calculus, including hypermuliplet matter. Section 3 reviews the orbifold compactification of the fifth dimension to an interval, the consequent breaking of higher-dimensional local supersymmetry down to 4D N=1 supersymmetry at the two boundaries as well as in the low energy 4D effective theory. Section 4 massages the bulk contribution to the effective tree-level potential for 4D Poincare-invariant VEVs into a useful form. Section 5 discusses the possible couplings of bulk supergravity (including the compensators) and bulk matter to the boundaries. New couplings are reported, and care is given to the consistent couplings of the compensators in light of the conformal anomaly of the visible sector dynamics. This is the key to understanding AMSB in this 5D context. Section 6 discusses supersymmetric radius stabilization. Section 7 discusses the corrections to bulk fields arising from hidden sector SUSY breaking and the cancellation of the 4D effective cosmological constant. It also explicitly connects the stabilizing fields to visible AMSB. Section 8 studies the form of SUSY breaking transmitted from the hidden boundary to the visible sector, thereby establishing sequestering with negligible corrections. Section 9 discusses radiative corrections to visible SUSY breaking due to loops of bulk fields. Gravity loops are known to provide an attractive solution to the AMSB tachyonic slepton problem of the MSSM (for hiddens sectors with large D-term SUSY breaking), but we also estimate the dangerous non-flavor-blind effects of loops of stabilizing fields. In Section 10, we give sample numerical estimates to show that successful AMSB is achievable, in particular with a gravity-loop resolution of the tachyonic slepton problem and adequate suppression of dangerous stabilizing-field loops.  The appendices contain some useful notation and formulae.

\vspace{0.5cm}

Throughout the remainder of the paper, unless otherwise specified,
we will use 5D Planck units, $M_5 \equiv 1$.

\section{5D Bulk Superfields}
\label{5BS}
Rigid 5D superconformal symmetry consists of translational symmetry ${\bf P}_A$, Lorentz symmetry ${\bf M}_{AB}$, ordinary supersymmetry ${\bf Q}$, special conformal symmetry ${\bf K}_A$, special supersymmetry ${\bf S}$, and dilatation ${\bf D}$ and internal $SU(2)$ ${\bf U}$ symmetry. In the Superconformal Tensor Calculus approach to 5D supergravity~\cite{FujitaO:2001,Kugo56:2002,KugoMatter:2002}, these symmetries are gauged and corresponding gauge fields are
\footnote{Throughout this paper, we will use $A,B,...= \dot 0, \dot 1, \dot 2, \dot 3, \dot 5$ for the flat 5D spacetime indices and $M,N,... = 0,1,2,3,5$ for the curved 5D indices. Similarly we will use $a,b,...= \dot 0, ..., \dot 3$ for the flat 4D spacetime indices and $\mu,\nu,... = 0,...,3$ for the curved 5D indices.}
\begin{equation}
   e^A_M, \omega_M^{\ AB}, \psi_M, f^A_M, \phi_M, b_M, \vec V_M
\end{equation}
where $\omega_M^{\ AB}, f^A_M, \phi_M$ become dependent fields in terms of the other independent fields by application of constraints.  $e^A_M$ is the f$\ddot{\rm u}$nfbein and $\psi_M$ is the gravitino. In off-shell formalism the independent gauge fields along with auxiliary fields, $v^{AB}, \chi^i, {\cal C}$ , form 5D Weyl multiplet given by
\begin{equation}
   (e^A_M, \psi^i_M, V_M^{ij}, b_M, v^{AB}, \chi^i, {\cal C}).
\end{equation}
The system contains three more supermultiplets:
two types of 5D hypermultiplets in a compact notation,
\begin{equation}
\label{hyper}
  (A^{\alpha}_i, \zeta^\alpha, F^\alpha_i),
\end{equation}
where, $\alpha = 1,2$ correspond to the unphysical compensator while $\alpha =3,4$ denote the physical hypermultiplet, and a 5D central charge vector multiplet,
\begin{equation}
    (M, A_M, \Omega^i, Y^{ij}).
\end{equation}
The ``graviphoton" $A_M$ appears within a seperate supermultiplet from graviton field, $e^A_M$, but in the action it mixes with $v_{AB}$ which is in the Weyl multiplet.

The hypermultiplet compensator is needed in order to derive physically consistent superconformal gravity invariant action and is used to gauge fix $SU(2)$ ${\bf U}$ symmetry while central charge vector multiplet fixes ${\bf D}$ and ${\bf S}$ symmetries. After these gauge fixings the theory reduces to 5D Poincare supergravity. The $i$ denotes the $SU(2)$ index while the $\alpha$ denotes $Usp(2,2)$ index. As the above indices imply, we will treat the hypermultiplets as the complex quantities and impose the reality constraint, i.e. $ A^i_\alpha \equiv -A^{\alpha *}_i = \epsilon^{ij}A^\beta_j \rho_{\alpha\beta}$ for the scalars
(same reality condition for $F^{\alpha}_i$ terms) and $\bar\zeta^\alpha\equiv (\zeta_\alpha)^\dagger\gamma^0={\zeta^\alpha}^T{\bf C}$ for the fermions where ${\bf C}$ is the 5D charge conjugation matrix.

The $SU(2)$ tensor, e.g. $Y^{ij}$, is related to the isovector, $\vec Y = (Y^1, Y^2, Y^3)$, by the relation,
\begin{equation}
   Y^i_{\, j} = Y^{ik}\epsilon_{kj} = i \vec Y \cdot \vec \sigma^i_{\, j}\, ,
\end{equation}
and it satisfies hermiticity $Y^{ij}=(Y_{ij})^*$. 

The details of the supersymmetry transformations, the construction of the action and the gauge fixing can be found in Refs.~\cite{FujitaO:2001,Kugo56:2002,KugoMatter:2002}.

\section{Compactification on $S^1/Z_2$}
\label{CS1Z2}
\subsection{Parity Assignment}
Regarding the orbifolding on $S^1/Z_2$, we will preserve $T^3 = 1/2\, \sigma^3$ generater to break $SU(2)_R$ down to $U(1)_R$ and the parity assignment can be carried out consistently by the transformation law under the $Z_2$, i.e. for the bosonic and fermionic fields,
\begin{equation}
\label{z2}
\begin{split}
   A(x^5) &\rightarrow {\cal P}A(-x^5),\\
   \zeta(x^5) &\rightarrow {\cal P}i\,\sigma^3\gamma^{\dot 5}\zeta(-x^5),
\end{split}
\end{equation}
where the eigenvalues ${\cal P}= +1, -1$ correspond to the parity even, odd respectively~\cite{KugoOrbifolding:2002}.

We summarize the parity assignments of the fields. The parity even fields are given by
\begin{equation}
\label{even}
 {\cal P}=+1: (e^a_\mu, e^{\dot 5}_5, \psi_{\mu+}, \psi_{5-}, b_\mu, V^{1,2}_5, V^3_\mu, v^{a\dot 5}, \chi_+, {\cal C};\,M, A_5,\Omega _- ,Y^{1,2};\, A^{2\hat\alpha-1}_{i=1}, A^{2\hat\alpha}_{i=2}, \zeta^{\hat\alpha}_-, F^{2\hat\alpha-1}_{i=2} ,F^{2\hat\alpha}_{i=1} )
\end{equation}
while the odd fields are
\begin{equation}
\label{odd}
 {\cal P}=-1: (e^{\dot 5}_\mu, e^a_5, \psi_{\mu-}, \psi_{5+},b_5, V^{1,2}_\mu, V^3_5,v^{ab}, \chi_-;\, A_\mu, \Omega _+, Y^3\,;\, A^{2\hat\alpha-1}_{i=2}, A^{2\hat\alpha}_{i=1}, \zeta^{\hat\alpha}_+, F^{2\hat\alpha-1}_{i=1} ,F^{2\hat\alpha}_{i=2} ),
\end{equation}
where $\hat\alpha=1$ corresponds to the compensator hypermultiplet and $\hat\alpha=2$ corresponds to the physical hypermultiplet. The orbifolding breaks 5D supersymmetry down to 4D N=1 supersymmetry.

\subsection{Induced 4D Boundary SUSY Multiplets}
\label{subI4SMB}
The orbifolding breaks 5D superconformal gauge symmetry down to 4D superconformal gauge symmetry. After gauge fixing the extra superconformal symmetries this corresponds to breaking 5D local supersymmetry to N=1 4D local supersymmetry. The 4D superconformal symmetry constrains any boundary action terms and it is therefore important to determine how bulk fields decompose under this symmetry.

The consistent parity assignment uniquely determines the following 4D  superconformal Weyl ( or gravity) multiplets
\footnote{The definitions of the supercovariant curvatures such as $\hat R_{\mu\dot 5}(Q)$ in the right hand side in Eq. (\ref{4dweyldef}) are given in Ref.~\cite{KugoOrbifolding:2002}. The definition of $\Gamma _5$ is given in Appendix \ref{ac}.}
at the boundary (i.e. it satisfies local SUSY transformation law of 4D superconformal gravity)~\cite{KugoOrbifolding:2002},
\begin{equation}
\label{4dweyl}
   (e^a_\mu, \psi_{\mu+}, b_\mu, a_\mu, \omega_\mu^{\hspace{2mm}ab}, \hat \phi_\mu, \hat f^a_\mu), \
\end{equation}
where
\begin{equation}
\label{4dweyldef}
\begin{split}
  a_\mu &=\frac{4}{3}(V^3_\mu + v_{\mu\dot 5})\, , \\
  \hat \phi_\mu &= \phi_\mu - \Gamma_5\gamma^a v_{a\dot 5}\psi_{\mu+} + \frac{1}{2}i\Gamma_5\hat R_{\mu\dot 5}(Q)_-\, , \\
  \hat f^a_\mu &= f^a_\mu -\frac{i}{2}\bar\psi_{\mu+}\Gamma_5\hat R^{a}_{\ \dot 5}(Q)_- - \frac{1}{6}\epsilon_\mu^{\hspace{3mm}abc}(\hat {\cal D}_b v_{c\dot 5} + \frac{1}{2}\hat R_{bc}^{\hspace{3mm}\dot 5}(V) ) + \frac{1}{4}\hat R_{\mu\dot 5}^{\hspace{4mm}a\dot 5}(M).
\end{split}
\end{equation}
This boundary restricted multiplet corresponds to the gauge fields of 4D superconformal symmetry~\cite{KakuTN,Ferrara:1978rk,Kugo:1982cu,KugoU:1983}. 

The graviphoton $A_M$ can also couple at the boundary via a parity even real general type 4D multiplet made from its gauge invariant field strength~\cite{KugoOrbifolding:2002}.
\begin{equation}
\begin{split}
   {\bf {\cal V}} = & ( M, -2i\Gamma_5 \Omega_-, 2Y^1, 2Y^2, \hat F_{a\dot 5} + 2v_{a\dot 5}M, -2 \hat {\cal D}_{\dot 5}\Omega_+ + 2i\gamma^a v_{a\dot 5}\Omega_- - \frac{i}{4}\Gamma _5\chi_+ M ,\\
   & \hat{\cal D}^2_{\dot 5}M - 2 \hat{\cal D}_{\dot 5}Y^3 - \frac{1}{4}{\cal C} M + v^a_{\ \dot 5}(2 \hat F_{a\dot 5} + v_{a\dot 5} M) +\frac{1}{2}\bar\chi_+\Omega_- )\, .
\end{split}
\end{equation}
It has a superconformal weight $(w,n)=(1,0)$ and has the lowest component given by ``dilaton" $M$. Note that $w$, $n$ denote Weyl weights, chiral weights respectively and $n$ is related to the standard chiral weights (or $R$-charge), $R$, via $n = \frac{3}{2} R$.

Each 5D hypermultiplet is decomposed into a vector-like pair of N=1 chiral multiplets with weight $(w,n)=(3/2,3/2)$ under the $S^1/Z_2$ orbifolding~\cite{KugoOrbifolding:2002}, i.e.
\begin{equation}
\label{4dhypermul}
\begin{split}
 &{\cal P}=+1: \quad (A^{2\hat\alpha}_{i=2}, -2i{\cal P}_R\zeta^{2\hat\alpha}, iF^{2\hat\alpha}_{i=1} + \hat{\cal D}_{\dot 5}A^{2\hat\alpha}_{i=1} )\, , \\
 &{\cal P}=-1: \quad (A^{2\hat\alpha-1}_{i=2}, -2i{\cal P}_R\zeta^{2\hat\alpha-1}, iF^{2\hat\alpha-1}_{i=1} + \hat{\cal D}_{\dot 5}A^{2\hat\alpha-1}_{i=1} )\, .
\end{split}
\end{equation}

\subsection{SUSY Multiplets in the 4D Effective Theory}
\label{subSM4FT}
After compactification, most field modes get masses of order $1/r$, where $\pi r$ is the length of the extra dimension. The approximate zero-modes are $x_5$-independent (hence parity-even) fields. The most obvious of these are $C, H$ and the 4D Weyl (gravity) multiplet. These are the analogs of the boundary-induced fields in the subsection \ref{subI4SMB}, except for being $x_5$-independent as opposed to restricted to the boundaries. The 4D Weyl multiplet zero-mode forms the gauge fields of the preserved 4D superconformal symmetry of the effective field theory below the compactification scale.

In addition, we should include boundary-localized fields (to be discussed) as well as a 4D radion multiplet. One can form a radion chiral multiplet incorporating the zero-modes of $e_5^{\dot 5}$ and $A_5$~\cite{KugoOrbifolding:2002},

\begin{equation}
\begin{split}
 -2i{\bf \Sigma}  = & \ (e^{\dot 5}_5 M - i A_5, -4i\, \psi_{5-} M + 4e^{\dot 5}_5 \Gamma_5 \Omega_-, -2i(V^1_5 + i V^2_5)M \\ 
  &- 2e^{\dot 5}_5 (Y^1 + i Y^2) + 2i \psi_{5-}(1+ \Gamma_5)\Omega_-).
\end{split}
\end{equation}
The 4D effective theory also has the zero-mode  $e_5^{\dot 5}$ appearing within a 4D radion general multiplet~\cite{KugoOrbifolding:2002},
\begin{equation}
\begin{split}
   {\bf W} = &\ (e^{\dot 5}_5, -2i \psi_{5-}, - 2V^2_5, 2 V^1_5, -2v_{a5},\\
    &\frac{i}{4}\Gamma_5 \chi_+ e^{\dot 5}_5 + 2\phi_{5+} + 2\Gamma_5 \gamma^b v_{b\dot 5} \psi_{5-}, (\frac{1}{4}{\cal C} - (v_{a\dot 5})^2)e^{\dot 5}_5 - e f^{\dot 5}_5 + \frac{i}{4}\bar\chi_+\Gamma_5 \psi_{5-})\, .
\end{split}
\end{equation}
In general, both these forms of radion field are required to write the locally supersymmetric 4D effective Lagrangian.

\section{Bulk Potential for 4D Scalars}
\label{BP4S}
In order to determine the supergravity and stabilization background in which the visible sector fields propagate we must determine the VEVs of all fields in the 4D Poincare invariant vacuum. This requires us to minimize the classical potential for all 4D scalars, even when these scalars arise from extra-dimensional components of 5D vectors and tensors. In this Section we summarize the contribution to this potential from 5D bulk action~\cite{KugoMatter:2002}. The most general 4D Poincare invariant geometry is given by the warped metric
\begin{equation}
     ds^2 = e^{2\sigma(\varphi)}\eta_{\mu\nu}dx^\mu dx^\nu - r^2 d\varphi^2\, .
\end{equation}
We are taking the fifth dimensional coordinate to be the ``angle" $\varphi$. While we do not wish to consider a highly warped spacetime, a mild warp factor will necessarily arise below as a backreaction to the stabilizing fields.

The effective potential from plugging in the 4D Poincare invariant ansatz into the bulk action can be economically written in compact flat (global) 4D superspace notation (See Appendix \ref{BP4SSN} for the derivation. See Refs.~\cite{CorreiaST:2005,AbeS:2004sd,CorreiaSTT:2006} for related work
\footnote{See Refs.~\cite{ArkaniHamed:2001tb,Marti:2001iw} for some early development of using 4D superspace for 5D action.})
\begin{equation}
\label{superbulkpoten}
\begin{split}
 - V_{bulk} = -\int_0^\pi  d\varphi & \left \{ \int d^4\theta\,e^{2\sigma}\frac{1}{2} (T+T^*) \left (2CC^* + 2\bar C\bar C^* - H^* e^{\frac{2}{\sqrt{3}}m\, \tilde V} H - \bar H e^{-\frac{2}{\sqrt{3}} m\, \tilde V}\bar H^* + \frac{1}{2} \tilde {\cal V}^3 \right ) \right . \\
   & + \int d^2 \theta\,e^{3\sigma}\,\left ( H(\frac{1}{2}\, \stackrel{\leftrightarrow}{\partial _5} +\, 2i\, m\Sigma)\bar H - C \stackrel{\leftrightarrow}{\partial _5} \bar C + \frac{i}{2}\, \Sigma\, \tilde{\cal W}_\alpha \tilde {\cal W}^\alpha + {\rm  h.c.} \right )\\
   &+ e^{4\sigma}r\, \left \{ -6\,M |Y|^2 +  V^3_5 V^{3\, 5} {\cal A}^2 \right . \\
   &\hspace{1.5cm} \left . + (\, 2\,m\,H\bar H Y + i V_5^3 (\, H^*\nabla ^5 H + \bar H^*\nabla^5\bar H  - 2\, C^* \partial ^5C - 2\,\bar C^* \partial^5 \bar C\, ) + {\rm h.c.} )\right \}\\
   & \left . + e^{4\sigma}r\,{\cal C}'({\cal A}^2 + 2M^3)\, \right \} ,
\end{split}
\end{equation}
in terms of the ``fake" flat superspace multiplets,
\begin{equation}
\begin{split}
\label{superbulkpotenDef}
  \tilde V &= \frac{1}{2}\theta^2\bar\theta^2(\sqrt{3}\, e^{2\sigma} D)\, , \quad \tilde{\cal W}_\alpha = \theta_\alpha\, e^{\sigma/2}\,\sqrt{3}\, D\, , \\
  \quad \tilde {\cal V} &= M - \theta^2\bar\theta^2  \partial_{\dot 5} ( e^{2\sigma}\,D)\, ,\\
   T &= \frac{-2i}{M}\Sigma = T  + \theta^2 e^\sigma {\cal F}_T\, , \quad T \equiv r -\frac{i}{M}A_5\, ,\\
   C &= C + \theta^2 e^\sigma {\cal F}_C \, ,\quad  \bar C = \bar C + \theta^2 e^\sigma {\cal F}_{\bar C}\, , \quad 
   H = H + \theta^2 e^\sigma {\cal F}_H\, , \quad \bar H = \bar H + \theta^2 e^\sigma {\cal F}_{\bar H}\, , \\
  {\cal A}^2 &= 2(-|C|^2 - |\bar C|^2 + 1/2\, |H|^2 + 1/2\, |\bar H|^2)\, .
\end{split}
\end{equation}
These fields are all functions of $\varphi$. Note that $H$,$\bar H$,$C$,$\bar C$ in the above equations are convenient renaming of the 4D N=1 hypermultiplets given by Eq. (\ref{4dhypermul}): the scalar components of these fields are related to $A^{2 \hat\alpha}_{i=2}$, $A^{2 \hat\alpha-1}_{i=2}$ in Eq. (\ref{4dhypermul}), e.g. $H \equiv A^4_2$, $\bar H \equiv A^3_2$, $C \equiv A^{2}_2$, and $\bar C \equiv A^1_2$ (Similarly for the $F$ terms of Eq. (\ref{4dhypermul})). Similarly $D$ in the first and second equations of Eq. (\ref{superbulkpotenDef}) is covenient renaming of $Y^3$ (See Eq. (\ref{appMdef}) in Appendix \ref{BP4SSN}).
The ${\cal F}$ terms and $D$ term, as functions of the original 5D $F$ terms and other 5D fields, are given in Appendix \ref{BP4SSN}. We will not require these expressions for our analysis. While the straight $F$ denotes F-term appearing in the 5D Lagrangian the curly ${\cal F}$'s denote those of the 4D chiral supermultiplets.

The covariant derivative in Eq. (\ref{superbulkpoten}) is defined as
\begin{equation}
\label{covHHbar}
   \nabla _5 H =  (\partial _5 - im A_5)H\, ,\quad \nabla _5 \bar H =  (\partial _5 + im A_5)\bar H\, .
\end{equation}
The ${\cal C}'$ is simply a Lagrange multiplier which forces the following contraint of the 5D superconformal gravity (See Ref.~\cite{KugoMatter:2002} for the relation between $\cal C$ and ${\cal C}'$),
\begin{equation}
\label{forcingconst}
  M^3 = |C|^2 + |\bar C|^2 - \frac{1}{2}\, |H|^2 - \frac{1}{2}\, |\bar H|^2\, .
\end{equation}
Note that in 5D supergravity a hypermultiplet mass $m$ must also be hypermultiplet charge for the central charge gauge boson, $A_M$.
Also note that under parity $\partial _5$ is odd while $A_5$ is even. Therefore, in the orbifolded theory we are forced to consider the hypermultiplet mass to be $\varphi$-dependent, $m = m_0\, \epsilon (\varphi)$, where $\epsilon(\varphi)$ is equals to $+1$ for $\varphi > 0$ and $-1$ for $\varphi < 0$.

It is convenient to write in the flat superspace notation because all the $\theta$-dependent components of Eq. (\ref{superbulkpotenDef}) as well as $V_5^3$ and $Y \equiv Y^1+ iY^2$ form the 4D N=1 supersymmetry order parameters. In particular, their vanishing (and the vanishing of any boundary auxiliary fields) in a classical solution guarantees the preservation of 4D supersymmetry. Note that 5D auxiliary fields such as $F$ are order parameters of the 5D supersymemtry not the 4D subalgebra. These are less informative since orbifolding has already broken 5D supersymmetry.

Note that the bulk potential given by Eq. (\ref{superbulkpoten}) has the explicit dependence of $M$ and $\bar C$ which means that the scale invariance and $SU(2)$ ${\bf U}$ symmetry have not yet been gauge-fixed.

\section{Boundary Physics}
\label{BP}
Boundary action terms invariant under the N=1 4D local supersymmetry are formed straightforwardly by writing 4D superconformal invariants in the standard fashion, but using the boundary-induced 4D gravitational fields.

\subsection{The Visible Sector Classical Action}
\label{subVSCA}
The visible action to be added to the bulk action takes the form
\begin{equation}
\begin{split}
  {\cal S }_{vis} &= \int d^5x \, \frac{\delta(\varphi -\pi)}{r}\,  {\cal L}_{vis} \, .
\end{split}
\end{equation}
We consider the visible sector consisting of N=1 chiral superfields, $Q$ as well as some gauge supermultiplets $V$. These dynamical fields couple to the 4D Poincare invariant VEVs of the boundary-induced bulk fields.
\begin{equation}
\label{visgeneralcoup}
\begin{split}
  {\cal L}_{vis} =& \int d^4 \theta \,e^{2\sigma} \,(CC^*)^{2/3}\, f(Q, Q^*, V ,H', {H'}^*,\partial _5 \bar C',\partial _5 \bar {C'}^*, \partial _5 \bar H',\partial _5 \bar {H'}^*,\ {\cal V}')  \\
  & + \int d^2 \theta\,e^{3\sigma}\,( C^2\,W_{vis}(Q) +  J_\pi C H ) + {\mathrm h.c.} \\
  & + \int d^2 \theta\, \tau(Q) {\cal W}_\alpha {\cal W}^\alpha + {\rm h.c.}\, .
\end{split}
\end{equation}
We have explicitly written a superpotential for the hypermultipet fields to aid in radius stabilization. Note that we have chosen these protected couplings $W_{vis}$ and $\tau$ to be independent of the bulk fields, as is technically natural.

Here, we have written the most general K${\ddot {\rm a}}$hler term constrained to have superconformal weight $(2,0)$. For this purpose it has been convenient to define the primed combinations of the boundary-induced bulk fields with zero superconformal weight $(0,0)$,
\begin{equation}
\label{pf}
 {\cal V}\,' = \frac{{\cal V}}{(CC^* + \bar C\bar C^* -1/2\ HH^* - 1/2\ \bar H \bar H^*)^{1/3}}\, ,\quad (H', \bar H', \bar C' ) = \frac{(H,\bar H,\bar C)}{C} \, .
\end{equation}
Note that the $e^{\dot 5}_5$-dependent ``radion" multiplet ${\bf \Sigma}$ and ${\bf W}$ can not couple gauge invariantly to the boundary as explained in Ref.~\cite{KugoOrbifolding:2002}.

The visible action appears very similar to purely 4D action in superconformal gravity if one makes the field redefinition $C\rightarrow C^{2/3}$ in order to get the more familiar 4D compensator of weight $(1,1)$. However we will stick to our weight $(3/2,3/2)$ compensator definition inherited from the bulk theory.

\subsection{Sequestering}
\label{subS}
In Section \ref{4ETS} we will show the very important result that, after radius stabilization, the VEVs of all the primed fields in Eq. (\ref{pf}) have negligible 4D supersymmetry breaking components. That is, these VEVs are essentially pure numbers which merely renormalize the visible K${\ddot {\rm a}}$hler and superpotentials. Therefore, the visible sector fields effectively propagate in only an induced-compensator background,
\begin{equation}
\begin{split}
  {\cal L}_{vis} =& \int d^4 \theta \,e^{2\sigma}\, (CC^*)^{2/3}\, f(Q, Q^*, V) \\
  & + \int d^2 \theta\,e^{3\sigma}\, ( C^2\,W_{vis}(Q) +  J_\pi C H ) + {\mathrm h.c.} \\
  & + \int d^2 \theta\, \tau(Q) {\cal W}_\alpha {\cal W}^\alpha + {\rm h.c.}\, .
\end{split}
\end{equation}
In particular the only 4D supersymmetry breaking felt by the visible sector fields comes from ${\cal F}_C$. This feature is sequestering.

\subsection{Visible Quantum Dynamics and Anomaly Mediation}
The compensator-dependence of the visible action is dictated by the requirement of maintaining exact superconformal gauge symmetry. To focus on this aspect we will keep only the renormalizable terms of the visible sector fields.
\begin{equation}
\begin{split}
  {\cal L}_{vis} \approx& \int d^4 \theta \, e^{2\sigma}\,(CC^*)^{2/3}\, Q^*e^V Q \\
  & + \int d^2 \theta\,e^{3\sigma}\, C^2\, (m_Q  Q^2 + \lambda Q^3 ) + {\mathrm h.c.} \\
  & + \int d^2 \theta\, \tau {\cal W}_\alpha {\cal W}^\alpha + {\rm h.c.}\, .
\end{split}
\end{equation}
Now redefine $e^\sigma\, C^{2/3}\,Q \rightarrow Q$ to arrive at 
\begin{equation}
\begin{split}
  {\cal L}_{vis} \approx& \int d^4 \theta \, Q^*e^V Q \\
  & + \int d^2 \theta\, ( m_Q\, e^\sigma\, C^{2/3}Q^2 + \lambda Q^3 ) + {\mathrm h.c.} \\
  & + \int d^2 \theta \, \tau {\cal W}_\alpha {\cal W}^\alpha + {\rm h.c.}\, .
\end{split}
\end{equation}
In this form it is clear that $e^\sigma \, C^{2/3}$ compensates for the breaking of superconformal invariance due to visible masses.

However, at the quantum level the running of visible couplings also violates superconformal invariance and it is vital that $C$ also compensates for this, 
\begin{equation}
\begin{split}
  {\cal L}_{vis} \approx& \int d^4 \theta\, Z \left (\mu \, e^\sigma (CC^*)^{1/3} \right )\, Q^*e^V Q \\
  & + \int d^2 \theta\, ( m_Q\, e^\sigma  C^{2/3}Q^2 + \lambda Q^3 ) + {\mathrm h.c.} \\
  & + \int d^2 \theta \, \tau_{1-loop}(\mu \, e^\sigma C^{2/3}) {\cal W}_\alpha {\cal W}^\alpha + {\rm h.c.}\, .
\end{split}
\end{equation}
As reviewed in Introduction this leads to the pattern of anomaly mediated supersymmetry breaking in the visible sector, seeded by ${\cal F}_C$.

This demonstration of anomaly mediation on the visible boundary directly in the 5D set-up (once we demonstrate the suppression of SUSY breaking within the primed bulk fields, as done in Section \ref{4ETS}), is one of the main results of this paper.

\subsection{The Hidden Sector and SUSY Breaking}
\label{subHSSB}
The hidden sector action takes the form 
\begin{equation}
\begin{split}
  {\cal S }_{hid} &= \int d^5x \, \frac{\delta(\varphi)}{r}\,  {\cal L}_{hid} \, ,
\end{split}
\end{equation}
where ${\cal L}_{hid}$ takes the entirely analogous form to ${\cal L}_{vis}$ with the visible sector fields replaced by some hidden sector fields responsible for supersymmetry breaking at the intermediate scale $\Lambda \sim 10^{11.5}$ GeV. The warp factor $e^{\sigma(\varphi)}$ can be taken to be unity at any point in the extra-dimensional interval. We will choose this location to be at the hidden boundary, $\sigma(\varphi =0)=0$. We assume that all hidden sector VEVs (whether SUSY breaking or preserving) are at most of order $\Lambda$ to the appropriate power, and all hidden masses are of order $\Lambda$ or above, except for a massless Goldstino eaten by the 4D effective gravitino.

The bulk couplings are only a slight perturbation to the massive hidden sector dynamics. Therefore the hidden VEVs are effectively given quantities that we can treat as sources for the bulk fields via their coupling in ${\cal S }_{hid}$. At this level, the dominant contributions to the effective potential for 4D scalars is given by
\begin{equation}
\label{hidpoten}
\begin{split}
 - V_{hid} &= - \Lambda^4 + \int d^2\theta\, ( c\,  C^2 + J_0 C H) + {\rm h.c.}\, .
\end{split}
\end{equation}
Again the hypermultiplet superpotential will aid in radius stabilization, while the constant superpotential with coefficient $c$ will be adjusted to cancel the 4D effective cosmological constant when supergravity effects are taken into account. Predominantly, this cancellation is against the positive supersymmetry breaking $\Lambda^4$ vacuum energy. Therefore $c \sim \Lambda ^2 M_4$, where $M_4$ is the effective 4D Planck scale.

There are also a variety of other couplings of hidden VEVs to bulk fields possible in ${\cal S }_{hid}$, essentially $\Lambda$-scale or smaller sources for bulk fields. These will have far weaker effects than the other sources, such as $J$, that we consider. However, we will have to consider such couplings more carefully when we are discussing the delicate issue of sequestering and visible SUSY breaking, in Section \ref{4ETS}. We will find that there too they play a negligible role.

\section{Supersymmetric Radius Stabilization}
\label{SRS}
We will study the supersymmetric radius stabilization introduced in Ref.~\cite{MaruO:2004}, and studied in the formalism of 5D superconformal tensor calculus in Ref.~\cite{AbeS:2004dr}. However our treatment will be somewhat different. Natural expansion parameters in our perturbative setup for the supersymmetric radius stabilization are $J_\pi$ and 5D hypermultiplet mass $m_0$. We are considering $J_\pi$ greater than $J_0$. We are dropping the constant superpotential $c$ which is a supersymmetric coupling, but the 4D cosmological constant cancellation will relate it to SUSY breaking $\Lambda^4$, and so we only include it when considering supersymmetry breaking in Section \ref{SB}.

The stabilization mechanism of radius modulus determines the size of extra-dimension in terms of these 5D parameters, i.e.
\begin{equation}
\label{sizeXD}
   \frac{1}{\pi r} \sim m_0 \, \left ( {\ln} \frac{J_\pi}{J_0} \right )^{-1} \, ,
\end{equation}
as will be discussed in detail in subsection \ref{subSBSS}. The 4D Planck mass, $M_4$, is also uniquely related to the size of extra-dimension by the relation (Recall that we are working in a 5D Planck unit, $M_5\equiv 1$.),
\begin{equation}
   M^2_4 = \pi r\, .
\end{equation}
The canonical mass of the radius modulus below the compactification scale is determined to be roughly
\begin{equation}
      m^2_{radion} \sim \frac{m_0^3\pi r\, |J_0|^2}{1 - e^{-2m_0\pi r}} \, .
\end{equation}
This will be discussed in subsection \ref{subRM}.

\subsection{Stabilization and Preserved SUSY}
\label{subSBSS}
The stabilization of radius modulus is carried out by the couplings of $H$ to the boundaries as well as 5D mass of the hypermultiplet as was indicated in Eq. (\ref{sizeXD})~\cite{MaruO:2004,AbeS:2004dr}.
\begin{equation}
\label{effpot1}
 - V_{eff} = - \int d \varphi\, \{ V_{bulk} + \int d^2\theta\, e^{3\sigma}\left (  \delta(\varphi)J_0 CH - \delta(\varphi-\pi)J_\pi CH + {\rm h.c.} \right )  \}\, .
\end{equation}
That is, in this Section we are neglecting visible sector, the SUSY breaking hidden sector, and the constant hidden superpotential, $c$, whose effect is to cancel the 4D SUSY breaking vacuum energy. The form of $V_{bulk}$ is given by Eq. (\ref{superbulkpoten}).

After performing superspace integration of the above full effective potential, the form in terms of 4D auxiliary fields is given by
\begin{equation}
\label{auxpot}
\begin{split}
  -V_{eff} = \int d\varphi\, e^{4\sigma} &\left \{ -\frac{3}{2} r D^2 - 3 D (\partial _5 \sigma) + mr D (|H|^2-|\bar H|^2) \right . \\
  &  -2r |{\cal F}_C|^2  - 2r |{\cal F}^*_{\bar C}|^2 + r |{\cal F}_H|^2 + r |{\cal F}_{\bar H} |^2  -6 r|Y|^2 - 2r V^3_5 V^{3\, 5} \\
  &  + 2{\cal F}_C  \{ -\frac{1}{2}C^* {\cal F}_T^* - (\partial _5 +\frac{3}{2} \partial _5\, \sigma \,)\bar C -  \frac{1}{2}J_0 H \delta(\varphi) + \frac{1}{2} J_\pi H \delta(\varphi -\pi) \} + {\rm h.c.}  \\
   &+ 2{\cal F}_{\bar C}\{ -\frac{1}{2}\bar C^* {\cal F}_T^* +  (\partial _5  + \frac{3}{2} \partial _5\, \sigma )C \} +{\rm h.c.}\\
  &+{\cal F}_H \{\, \frac{1}{2}{\cal F}^*_TH^* + (\partial _5  +\frac{3}{2}\partial _5 \sigma  - mT)\bar H   -J_0 C \delta(\varphi)+ J_\pi C \delta(\varphi -\pi) \}+ {\rm h.c.}\\
  &+{\cal F}_{\bar H} \{\,  \frac{1}{2}{\cal F}^*_T \bar H^* - (\partial _5  +\frac{3}{2}\partial _5 \sigma  + mT) H  \}+ {\rm h.c.} \\
  & -mH\bar H {\cal F}_T + {\rm h.c.} \\
  &\left . +2\,mr\,H\bar H Y + i\, r V_5^3 (\, H^*\nabla ^5 H + \bar H^*\nabla^5\bar H - 2\, C^* \partial ^5C - 2\,\bar C^* \partial^5 \bar C\, \, ) + {\rm h.c.} \right \} \, .
\end{split}
\end{equation}
As was discussed in Section \ref{BP4S}, all the $\theta$-dependent components of Eq. (\ref{superbulkpotenDef}) as well as $V^3_5$ and $Y$ form the 4D N=1 supersymmetry order parameters. Therefore, when these supersymmetry order parameters vanish the corresponding field configuration preserves 4D N=1 supersymmetry. These supersymmetry order parameters are given in terms of other fields by their equations of motion from Eq. (\ref{auxpot}),
\begin{equation}
\label{auxfields}
\begin{split}
   &{\cal F}^*_{\bar H} = \frac{1}{r} \{ (\partial _5 + \frac{3}{2}\, \partial _5 \sigma + mT)H - \frac{1}{2} {\cal F}^*_T \bar H ^* \}\, ,\\
   &{\cal F}^*_H = -\frac{1}{r}\{ (\partial _5 + \frac{3}{2}\, \partial _5 \sigma - mT)\bar H - J_0\, C\delta(\varphi) + J_\pi\, C \delta(\varphi-\pi)  + \frac{1}{2} {\cal F}^*_T H^* \} \, ,\\
   &{\cal F}^*_{\bar C} = \frac{1}{r} \{ (\partial _5  + \frac{3}{2}\, \partial _5 \sigma \,) C -\frac{1}{2} {\cal F}_T^* \bar C^* \}  \, ,\\
  &{\cal F}^*_C  = - \frac{1}{r} \{ (\partial _5  + \frac{3}{2}\, \partial _5\sigma\,) \bar C +  \frac{J_0}{2} H \delta(\varphi) - \frac{J_\pi}{2} H \delta(\varphi-\pi) + \frac{1}{2} {\cal F}^*_T  C^* \} \, ,\\
   &0 =-mH\bar H + \frac{1}{2}H {\cal F}^*_H + \frac{1}{2}\bar H {\cal F}^*_{\bar H} -\bar C {\cal F}^*_{\bar C} - C {\cal F}_C^*\, , \\
      &Y^* = \frac{1}{3}m H\bar H\, ,\\
   &D = \frac{1}{3}\, m (|H|^2-|\bar H|^2)- \frac{(\partial _5 \sigma)}{r}\, , \\
   &V_5^3 = \frac{i}{4} (H^*\nabla _5 H + \bar H^*\nabla _5 \bar H - 2\, C^* \partial _5 C - 2\,\bar C^* \partial_5 \bar C) + {\rm h.c.}\, ,
\end{split}
\end{equation}
where the definitions of $\nabla _5 H$, $\nabla _5 \bar H$ are given by Eq. (\ref{covHHbar}). We also integrate out ${\cal C}'$, appearing in Eq. (\ref{superbulkpoten}), resulting in the constraint
\begin{equation}
\label{constraint1}
\begin{split}
   |C|^2 &=  1 -|\bar C|^2 + \frac{1}{2}|H|^2 + \frac{1}{2}|\bar H|^2\, .
\end{split}
\end{equation}
In the remainder of this Section we will solve the set of equations,
\begin{equation}
\label{susyeom}
  {\cal F}_{H, \bar H, C, \bar C, T}= Y = D = V^3_5 =0\, ,
\end{equation}
that gives rise to 4D N=1 supersymmetric field configuration.
Note that the solutions of the set of equations of motion given by plugging Eq. (\ref{susyeom}) in Eq. (\ref{auxfields}) also satisfy the full set of second-order differential equations of the propagating fields as is clear in the form of Eq. (\ref{auxpot}). Further, we will seek a solution with $\bar C=0$.

The solution of the second equation of Eq. (\ref{auxfields}) is then given by
\footnote{The identity,
\begin{equation}
    (\epsilon(\varphi))^2 \delta(\varphi) = \frac{1}{3}\, \delta(\varphi)\, ,
\nonumber
\end{equation}
( similarly for $\delta(\varphi -\pi)$) is used to treat the  
$\epsilon(\varphi)$ functions inside $C$ given by Eq. (\ref{constraint1})~\cite{Bagger:2001s}.
}
\begin{equation}
\label{Hbarsol}
\begin{split}
   \bar H(x,\varphi) &= \epsilon(\varphi)\, \frac{J_0}{2}C(0)\, e^{m_0 T|\varphi| + \frac{3}{2}(\sigma(\varphi)-\sigma(0))}\, 
\end{split}
\end{equation}
and, to the leading order of $J$, to match the delta functions
\begin{equation}
\label{vevofT}
  J_0-J_\pi \,e^{-m_0 T \pi} \sim 0\, ,
\end{equation}
since to the leading order of $J_\pi$ we can neglect warp factor. $J_0, J_\pi$ can be made real by absorbing their phases into $H$, $C$. Therefore, the VEV of $T$ is just equal to $r$. Solving for $r$ using Eq. (\ref{vevofT}) gives us Eq. (\ref{sizeXD}). Taking into account that $\bar H$ has a non-vanishing field configuration, $Y=0$ equation of Eq. (\ref{auxfields}) implies
\begin{equation}
\label{Hvev}
    H(x,\varphi) =0\, .
\end{equation}
The first and fourth equations of Eq. (\ref{auxfields}) are now automatically satisfied. The warp factor $\sigma$ is determined by $D=0$ equation of Eq. (\ref{auxfields}) to be
\begin{equation}
\label{warpfactor}
  \partial _5 \sigma = - \frac{1}{6}m (T+T^*) |\bar H|^2\, .
\end{equation}
Note that the integration constant of $\sigma$ is determined by our convention, $\sigma(\varphi =0)=0$.
The third equation of Eq. (\ref{auxfields}) gives rise to
\begin{equation}
\label{Ceom}
   0 = \partial _5 |C| + \frac{3}{2}(\partial _5 \sigma )|C| + i (\partial _5 \vartheta)|C|\, ,
\end{equation}
where $C$ was parameterized as $C= |C|e^{i{\vartheta}}$. The imaginary part of the above equation gives rise to
\begin{equation}
\label{cphase}
 \partial _5 \vartheta =0\, .
\end{equation}
Using the constraint of $C$ given by Eq. (\ref{constraint1}) (with $\bar C$=0 as mentioned above) and the solution of $\bar H$ given by Eq. (\ref{Hbarsol}), the real part of Eq. (\ref{Ceom}) reduces to
\begin{equation}
\begin{split}
   0 &= \frac{1}{2}\partial _5 |C|^2 + \frac{3}{2}(\partial _5 \sigma)|C|^2 \\
     &=\frac{1}{4}(\,  m(T+T^*)|\bar H|^2 - 3(\partial _5 \sigma)|\bar H|^2 \, ) + \frac{3}{2}(\partial _5 \sigma )(1+ \frac{1}{2}|\bar H|^2)\, .
\end{split}
\end{equation}
In the second equality, we have used an identity, $(\epsilon(\varphi))^{2n+1}\delta(\varphi)=0$ ($n$= integer). The above equation is automatically satisfied by Eq. (\ref{warpfactor}). The last equation of Eq. (\ref{auxfields}) is also automatically satisfied by Eqs. (\ref{Hbarsol}), (\ref{Hvev}), (\ref{warpfactor}) and (\ref{cphase}).

In this way we have solved all of equations of Eq. (\ref{auxfields}) supersymmetrically, i.e. subject to Eq. (\ref{susyeom}).

Since the full effective potential is quadratic in the supersymmetry order parameters, i.e.
\begin{equation}
\begin{split}
  V_{eff} = \int d\varphi\, e^{4\sigma} &\left ( r|{\cal F}_H|^2 + r|{\cal F}_{\bar H}|^2 - 2r|{\cal F}_C|^2 -2 r|{\cal F}_{\bar C}|^2 -\frac{3}{2} r D^2 \right . \\
   &\left . + m H\bar H {\cal F}_T + m H^*\bar H^*{\cal F}_T^* - 6r |Y|^2 + \frac{2}{r}(V^3_5)^2 \, \right )\, ,
\end{split}
\end{equation}
the resulting vacuum energy on the locally supersymmetric solutions that satisfy the full set of equations in Eq. (\ref{susyeom}) identically vanishes,
\begin{equation}
   V_{eff} = 0\, .
\end{equation}

\subsection{Mass of Radion}
\label{subRM}
The mass of radion can be easily estimated from the 4D effective theory below the compactification scale. The 4D effective theory is described by the zero modes represented by $T$, $H_0$ and $C_0$ in this case while the other fields get masses of order $\sim\frac{1}{r}$ and they can be integrated out in a supersymmetric way. In order to derive the 4D effective theory of these light fields, we solve the equations of motion given by Eq. (\ref{auxfields}) supersymmetrically with $J_{0,\pi} = 0$ (In this limit these light fields become massless), e.g.
\begin{equation}
\label{rigidsol}
\begin{split}
   H(x,\varphi) = H_0(x)\, e^{-m_0 T |\varphi|}\, .
\end{split}
\end{equation}
After plugging in Eq. (\ref{rigidsol}) as well as $C_0$, $T$ as supermultiplets, including their ${\cal F}$ terms, into Eq. (\ref{effpot1}) and performing the $\varphi$-integration, the 4D effective theory for $T$, $H_0$ and $C_0$, to the leading order of $J_\pi$, is given by
\begin{equation}
\label{4dsusyeffpot}
\begin{split}
   -V_{eff} \approx  &\int d^4 \theta \, -(T + T^*)|C_0|^2+ \frac{1}{2m_0}(1 - e^{-m_0 (T + T^*)})|H_0|^{2} \\
 &- \int d^2 \theta\, ( J_0- J_{\pi}e^{-m_0 T} ) H_0 C_0 + {\rm h.c.}\\
 &- 6 \pi r |Y|^2 + (\, m_0 r \pi J_0 H_0 Y + {\rm h.c.})\, .
\end{split}
\end{equation}
Here, $\pi$ from $\varphi-$integration was absorbed into $T$. Note that supersymmetrically the stabilization gives rise to a supersymmetric Dirac mass for $\delta T$ and $H_0$ as is clear in the 4D effective superpotential in Eq. (\ref{4dsusyeffpot}).  The kinetic term of radion are given by the well-known no-scale structure~\cite{noscale},
\begin{equation}
\begin{split}
   {\cal L}_{radion} = -\sqrt{-g}\,\pi r {\cal R}\, .
\end{split}
\end{equation}
At leading (zeroth) order in $J$, the hypermultiplets have a conserved hypermultiplet number, so that they can not correct the radion effective action at tree level. 

The radion has its kinetic term in the Einstein frame via a Weyl transformation, $g_{\mu\nu} =  \frac{M^2_4}{2\pi r}\, \tilde g_{\mu\nu}$, i.e.
\begin{equation}
\label{noscaleEin}
\begin{split}
 {\cal L}_{radion} &= \sqrt{-g}\, (\, - \pi r {\cal R}- V_{eff} ) \\ 
   &= \sqrt{-\tilde g}\, ( \,- \frac{M^2_4}{2} \tilde{\cal R} + \frac{3}{4} M^2_4 \frac{(\partial _\mu r)^2}{r^2} - \tilde V_{eff} )  \, .
\end{split}
\end{equation}
Expanding about supersymmetric VEVs, following from Eqs. (\ref{vevofT}) and (\ref{Hvev}), the Einstein-frame effective potential to quadratic order in the  fluctuations $\delta T$, $H_0$ is
\begin{equation}
\label{veffwoc}
\begin{split}
    \tilde V_{eff} &= \frac{M^4_4}{4\pi^2 r^2}\, V_{eff} \\
   &\approx \frac{M^4_4}{4\pi^2 r^2}\, \left ( \frac{11}{6}\pi m^2_0 r |J_0|^2 |H_0|^2 + \frac{2m_0^3|J_0|^2}{1- e^{-2\pi m_0 r}}|\delta T|^2 \right )\, .
\end{split}
\end{equation}
We can get the physical mass of the Dirac state by just getting the mass of the scalar radion $\delta T$. It does not mix with $H_0$ as can be seen in Eq. (\ref{veffwoc}). Given the form of the kinetic term of the radion in Eq. (\ref{noscaleEin}), the physical mass of the radion is given by,
\begin{equation}
\label{radionmass}
\begin{split}
  m^2_{radion}  \approx \frac{2}{3}\,\frac{ m_0^3\pi r |J_0|^2}{1-e^{-2\pi m_0 r}}\, .
\end{split}
\end{equation}
This recovers the results of Refs.~\cite{MaruO:2004,AbeS:2004dr} (in their unwarped limits).

This implies that the 4D effective theory, below the stabilization scale $\sim m_{radion}$, only contains $C_0$ (as well as the light fields of the visible and hidden sectors).

\section{The SUSY-Breaking Vacuum}
In order to discuss the supersymmetry breaking we add a constant superpotential $c$ to our effective potential given by Eq. (\ref{effpot1}). Therefore, the natural expansion parameters in our perturbative setup now are $J_\pi$, $m_0$ and the constant superpotential $c$. The constant superpotential $c \ll J_\pi, m_0$ is related to the supersymmetry breaking vacuum energy $\sim \Lambda^4$ by the 4D cosmological constant cancellation,
\begin{equation}
\label{ctolambda}
    \Lambda^4 \sim \frac{|c|^2}{M^2_4}\, .
\end{equation}
In this sense the constant superpotential $c$ usefully parameterizes supersymmetry breaking. It induces a deviation from the supersymmetric VEVs.  E.g. the deviation of the VEV of $H_0$, the zero mode of $H$, is of order $c$,
\begin{equation}
  H_0 \sim \frac{c}{\pi m_0 r J_0}\, .
\end{equation}
This will be discussed in Section \ref{SB}.

The ``seed" of AMSB, ${\cal F}_C/C$, is given by the VEV of $H_0$ in the bulk from the 5D perspective,
\begin{equation}
   \frac{{\cal F}_C}{C} \sim - (\epsilon(\varphi))^2\, \frac{1}{2}m_0\,J_0^* H_0^*\, ,
\end{equation}
as will be discussed in Section \ref{subRCS}.

\subsection{Supersymmetry Breaking}
\label{SB}
We now consider the effects of the constant superpotential $c$ introduced in Eq. (\ref{hidpoten}) (and $\Lambda^4$ which ensures that we are correctly expanding about a 4D Poincare invariant vacuum).

The approximate 4D effective potential of the zero modes of the light fields $T$, $H_0$ and $C_0$ is now given by
\begin{equation}
\begin{split}
  -V_{eff} \approx  &\int d^4 \theta \, -(T + T^*)|C_0|^2  + \frac{1}{2m_0}(1 - e^{-m_0 (T + T^*)})|H_0|^{2}\\
 &- \int d^2 \theta\, C_0 \{  ( J_0- J_{\pi}e^{-m_0 T} ) H_0 + c \} + {\rm h.c.}\\
 &- 6 \pi r |Y|^2 + (\, m_0 r \pi J_0 H_0 Y + {\rm h.c.}) \\
 &-\Lambda^4\, .
\end{split}
\end{equation}
This is just Eq. (\ref{4dsusyeffpot}) with the addition of $c$ and $\Lambda^4$.

Any deviation from the supersymmetric VEVs, discussed in Section \ref{SRS}, must be proportional to some power of $c$. Noting this, after performing superspace integration, the 4D effective potential to order $ c^2$ is given by
\begin{equation}
\label{4deffpot}
\begin{split}
   V_{eff}
   \approx &\frac{11}{6}\, \pi|J_0|^2 m_0^2 r\,|H_0|^2 - (\,2\,c^*J_0 m_0 H_0 + {\rm h.c.}) \\
   &+ \frac{2m_0}{1-e^{-2\pi m_0 \pi r}}|J_0|^2 m_0^2 |\delta T|^2 + \Lambda^4\, .
\end{split}
\end{equation}
Since to order $ c^2$ the $\delta T$ has only a quadratic term, at the minimum of the 4D effective potential of Eq. (\ref{4deffpot})
\begin{equation}
    \delta T \sim {\cal O}(c^2 )\, .
\end{equation}

Now, in order to analyze the remaining part of Eq. (\ref{4deffpot}), we parameterize $H_0$ as $H_0=|H_0|e^{i\theta}$, and minimize the 4D effective potential with respect to $\theta$:
\begin{equation}
\begin{split}
     V_{eff} &\approx \frac{11}{6}\, \pi|J_0|^2 m_0^2 r\,|H_0|^2 - 4|c||J_0| m_0 |H_0|+ \Lambda^4.
\end{split}
\end{equation}
Finally, minimizing the above effective potential with respect to $|H_0|$ gives the minimum,
\begin{equation}
\label{minpot}
   V_{min} \approx -\frac{24}{11}\, \frac{|c|^2}{\pi r} + \Lambda^4\, \quad  {\rm at}\quad
    H_0 \approx \frac{12}{11}\frac{c}{\pi m_0 r J_0}\, .
\end{equation}
The cancellation of the 4D cosmological constant relates the constant superpotential $c$ to the positive vacuum energy,
\begin{equation}
    \Lambda^4 \approx \frac{24}{11}\frac{|c|^2}{M^2_4}.
\end{equation}

\subsection{Relation between Compensator and Stabilization}
\label{subRCS}
After some algebra, using the first five equations in Eq. (\ref{auxfields}), one can obtain the expressions of ${\cal F}_{H,\bar H, C,\bar C, T}$ in terms of the non-auxiliary fields. In particular,
\begin{equation}
\label{seedamsb}
\begin{split}
  \frac{{\cal F}^*_C}{C^*} =& - m H\bar H - \frac{2|\bar C|^2- |H|^2-|\bar H|^2}{2r\,C^*} 
  \left \{ (\partial _5 + \frac{3}{2}\partial _5\,\sigma \,)\bar C + \frac{J_0}{2}H\delta (\varphi) - \frac{J_\pi}{2}\delta (\varphi -\pi)H \right \}\\
  &- \frac{H}{2r}\left \{ (\partial _5 + \frac{3}{2}\partial _5 \, \sigma - mT\, )\bar H - J_0 C\delta(\varphi) + J_\pi C\delta(\varphi -\pi) \right \}\\
  &+\frac{\bar H}{2r}(\partial _5 +\frac{3}{2}\partial _5\,\sigma + mT\, )H -\frac{\bar C}{r}(\partial _5 + \frac{3}{2}\partial _5\, \sigma )C\, .
\end{split}
\end{equation}
The constraint of the lowest component of $C$ is given by Eq. (\ref{constraint1}).

It is convenient to work with parity even fields $\bar h(x,\varphi)$, $\bar c(x,\varphi)$ by parameterizing as $\bar H(x,\varphi)\equiv \epsilon(\varphi)\bar h(x,\varphi)$, $\bar C(x,\varphi)\equiv \epsilon(\varphi)\bar c(x,\varphi)$.  The boundary values of $\bar h$, $\bar c$ are determined by their equations of motion (See Ref.~\cite{Eto:2004yk} for related discussion), e.g. the singular parts of the equations of motion of $\bar H$, $\bar C$ from the Eq. (\ref{auxpot}) imply
\begin{equation}
\begin{split}
  0 &= - \frac{e^{4\sigma}}{r} \partial _5  \{ 2(\delta (\varphi) -\delta(\varphi -\pi) )\bar h - J_0 C \delta (\varphi) + J_\pi C \delta (\varphi-\pi) \}\, , \\
  0 &= - 2\frac{e^{4\sigma}}{r} \partial _5 \{ 2(\delta (\varphi) -\delta(\varphi -\pi) )\bar c + \frac{J_0}{2} H \delta (\varphi) - \frac{J_\pi}{2} H \delta (\varphi-\pi) \}\, .
\end{split}
\end{equation}
Solving the above equations gives rise to
\begin{equation}
\label{cbarbdy}
\begin{split}
   &\bar c(x,0) = - \frac{J_0}{4}H(x,0)\, , \quad \bar c(x,\pi) = -\frac{J_\pi}{4}H(x,\pi)\, ,\\
   &\bar h(x,0) = \frac{J_0}{2}C(x,0)\, , \quad \bar h(x,\pi)= \frac{J_\pi}{2}C(x,\pi)\, .
\end{split}
\end{equation}
These boundary values also remove all the singularities inside ${\cal F}_C$, ${\cal F}_H$ (therefore, singularities inside ${\cal F}_T $ as is indicated by the fourth equation of Eq. (\ref{auxfields}))

Taking into account that the VEV of $H_0 \sim {\cal O}(c)$ ( See  Eq. (\ref{minpot})), $\bar c(x,\varphi)$ is order $c$ by Eq. (\ref{cbarbdy}).  Note that unlike the supersymmetric case in which the VEV of $H_0$ vanishes, the parity odd $\bar C$ becomes discontinuous at the boundary.

Given the profiles of $H$, $\bar H$ and the warp factor $\sigma$ given by Eq. (\ref{warpfactor}), ${\cal F}_C/C$ to the leading order of our expansion parameters is approximately given by
\begin{equation}
\label{amsbseed}
   \frac{{\cal F}_C}{C} \sim - m H^*\bar H^* \sim - (\epsilon(\varphi))^2\, \frac{1}{2}m_0\,J_0^* H_0^*\, .
\end{equation}
A similar result was derived in Ref.~\cite{MaruO:2004} from 4D EFT viewpoint.
The above expression implies that the ``seed" of AMSB, ${\cal F}_C/C$, gets a non-trivial contribution only via the VEV of a stabilizing field in the bulk. Using the result given by Eq. (\ref{minpot}), the 5 dimensional ${\cal F}_C/C$ is given by
\footnote{Recall that $(\epsilon(\varphi))^{2n}\, \delta(\varphi)$ ($n$=integer) gives rise to a non-vanishing contribution while $(\epsilon(\varphi))^{2n+1}\, \delta(\varphi) =0$.}
\begin{equation}
\begin{split}
  \frac{{\cal F}_C}{C} &= -(\epsilon(\varphi))^2\,\frac{c^*}{2\pi r}\\
   &=-(\epsilon(\varphi))^2\,\frac{c^*}{2}\, m_0 \left ( {\rm ln} \frac{J_0^*}{J_\pi^*}\right )^{-1}\, .
\end{split}
\end{equation}
Eq. (\ref{amsbseed}) shows us that on-shell it is the stabilizing fields that are communicating SUSY breaking to the visible sector. Since these are not gravitational or gauge fields one might have expected their couplings to visible fields to be unconstrained enough that a rather general pattern of visible SUSY breaking would result. It is remarkable therefore that the SUSY breaking is communicated dominately via Eq. (\ref{amsbseed}) and its highly constrained couplings to the visible sector, resulting in the very specical pattern of AMSB. That other forms of SUSY breaking coupling are subdominant is argued below. This is the content of sequestering.

\section{4D Effective Theory and Sequestering}
\label{4ETS}
Here, we demonstrate sequestering at tree level, namely the primed bulk fields induced on the visible boundary have only supersymmetric VEVs, with negligible corrections, leading to the dominance of AMSB as discussed in Section \ref{BP}. This has not been previously done. We would like ideally to do this in a direct 5D analysis for maximal transparency but have not managed to do this, and so we instead we will partly exploit the organizing power of the 4D effective field theory below the compactification and stabilization scales. We are considering in this paper a hierarchy of scales,
\begin{equation}
1/r \gg m_{radion} \gg \Lambda,
\end{equation}
and will later show samples of parameter inputs that achieve this. Therefore it makes sense to study the effects of SUSY breaking within a 4D effective field theory below the radion stabilization scale. Below this scale the only light 4D multiplets are the visible fields, the hidden sector  fields, and 4D effective supergravity fields, including in particular the compensator zero-mode, $C_0$. If sequestering fails, that is if the prime fields in Section \ref{BP} acquire non-negligible SUSY breaking VEVs then clearly, by choice of couplings to the visible fields in Eq. (\ref{visgeneralcoup}), they can induce new visible soft terms. This visible SUSY breaking must be accounted for in the 4D EFT by visible couplings to $C_0$ and/or hidden fields, constrained by the effective 4D superconformal tensor calculus. We will show that this cannot happen without great suppression.

Let us begin by considering couplings which do involve hidden fields. The danger is in K$\ddot{\rm a}$hler terms of the form
\begin{equation}
\label{viscoupling}
\int d^4 \theta\, Q^* e^V Q\,  X, 
\end{equation}
where $X$  is a composite of hidden sector fields. For simplicity, we assign $X$ weight $(2,0)$ so the compensator is unnecessary. Hidden SUSY breaking VEVs in $X$ will certainly contribute to visible SUSY breaking, and spoil sequestering, if such a coupling arises from matching to 5D. But does it? Since the hidden and visible fields are spatially separated in 5D, effective couplings like the one above can only arise through integrating out some bulk fields that couple to both boundaries. Focussing on the hidden boundary, the leading contributions at tree-level (we deal with loops in Section \ref{BRC}) arise from K$\ddot{\rm a}$hler couplings of $X$ to bulk fields in ${\cal S}_{hid}$. At tree level precisely one of these bulk fields contracts with a Feynman diagram that ends on the visible sector, while the remaining bulk fields are set to their supersymmetric VEVs. We will use the primed basis for the 4D induced bulk superfields defined in subsection \ref{subVSCA}, and put a prefix ``$\delta$'' to indicate the one which connects to the visible sector. Then the general set of allowed K$\ddot{\rm a}$hler couplings are of the form, 
\begin{eqnarray}
\label{chnonch}
&~& \int d^4 \theta\, X\ \delta C \times {\rm SUSY~VEVs}\, ,  
\int d^4 \theta\, X\ \delta \partial_5 \bar{C}' \times {\rm SUSY~VEVs}\, ,
\int d^4 \theta\, X\ \delta H' \times {\rm SUSY~VEVs},  \nonumber \\
&~& \int d^4 \theta\, X\ \delta \partial_5 \bar{H}' \times {\rm SUSY~VEVs}\, ,
\int d^4 \theta\, X\ \delta {\cal V}' \times {\rm SUSY~VEVs}. 
\end{eqnarray} 
Since the SUSY VEVs are pure numbers they can be subsumed into the arbitrary 
coefficients of these couplings. 

Let us look at the special features of such terms. Since boundary restricted $\delta C$, $\delta\bar C'$, $\delta H'$, $\delta\bar H'$ are 4D chiral multiplets and since we are only interested in the tree level potential, we can drop $x_\mu$-dependence so that the $\int d^2\bar\theta$ only applies to $X$, giving a chiral result. That is, without loss of generality we could just as well have considered $X$ to be chiral in these cases, and coupled as
\begin{equation}
 \int d^2\theta\, X\ \delta C\, , \ \int d^2\theta\, X\ \delta\partial _5 \bar C'\, , \ \int d^2\theta\, X\ \delta H'\, , \ \int d^2\theta\, X\ \delta \partial _5\bar H'\, .
\end{equation}
The soft visible mass-squareds from Eq. (\ref{viscoupling}) are then order $|{\cal F}_X|^2$. If the hidden sector does not contain hidden gauge symmetry singlets, then $X$ is necessarily a composite field and ${\cal F}_X$ has dimension at least 3, ${\cal F}_X \sim {\cal O}(\Lambda^3)$. The soft visible mass-squareds will therefore be order $\Lambda^6$.

If $X$ is non-chiral in Eq. (\ref{viscoupling}), then there will be soft visible mass-squared of order $D_X$. Since non-chiral $X$ gauge invariants begin at dimension 2, $D_X$ can have dimension 4, $D_X \sim {\cal O}(\Lambda^4)$. This is the ``going rate" for worrying about non-sequestering. Indeed $\delta {\cal V}'$ is non-chiral and one might worry that it couples to $D_X$ in Eq. (\ref{chnonch}). But this coupling is $D_X\ \delta M'$ where the lowest component of ${\cal V}'$ is given by
\begin{equation}
 M' = \frac{M}{(|C|^2 + |\bar{C}|^2 - 1/2\, |H|^2 - 1/2\, |\bar{H}|^2)^{1/3}}\, .
\end{equation}
By the constraint of Eq. (\ref{forcingconst}), $M' \equiv 1$, $\delta M'=0$, so $D_X$ is not mediated to the visible sector. It is very important to note that if we had had further bulk vector multiplets beyond the ``graviphoton'' multiplet, then we can really mediate visible couplings to non-chiral $X$, because the scalar components of such vector multiplets are not constrained like $M'$. Thus we have a ``special bulk content'' in this sense, although we are within our rights to choose the light particle content of the bulk as effective field theorists. But a string theory construction along these lines would have to check this precondition. See Ref.~\cite{Anisimov:2001zz,Kachru:2007s} for related discussion of this point.

Since the cutoff of 4D EFT is $m_{radion}$ the suppression of visible soft mass-squareds that we have demonstrated, by $\Lambda^2$ compared to going rate corresponds to a dimensionless suppression of at least $\Lambda^2/m_{radion}^2$. In this paper, we take the simple approach of ensuring that $\Lambda^2/m_{radion}^2$ is very small, numerically of order $10^{-7}$, so as to adequately suppress its contributions compared with AMSB by enough so to retain the attractive AMSB solution to the supersymmetric flavor problem.

We have shown that we can neglect hidden sector fields in checking sequestering, but this still leaves effective visible-compensator couplings to check. By 4D superconformal gauge symmetry of low energy supergravity in the superconformal tensor calculus, the leading tree level couplings of the visible sector to a Poincare invariant VEV of the 4D supergravity fields, must simply be given by
\begin{equation}
{\cal L}_{eff} = \int d^4 \theta\, Q^{\dagger} e^V Q + \int d^2 \theta\, 
W(Q,C_0) +
\int d^2 \theta\, \frac{1}{g^2} {\cal W}_{\alpha}^2
+{\rm h.c.},
\end{equation}
where the superpotential is exactly cubic in fields to maintain classical conformal invariance. This form manifests the compensating role of $C_0$ in the effective theory. Here $Q$ has unit superconformal weight. The corresponding more standard form with zero weight for $Q$ is given by
\begin{equation}
{\cal L}_{eff} = \int d^4 \theta\, |C_0|^2
Q^{\dagger} e^V Q + \int d^2 \theta\, C_0^3 W(Q) +
\int d^2 \theta\, \frac{1}{g^2} {\cal W}_{\alpha}^2
+ {\rm h.c.},
\end{equation}
where the new $Q$ is  $Q/C_0$ in terms of the old. Note that $C_0$ is the standard version of the effective 4D compensator with superconformal weight $(1,1)$.

We see that any classically conformally invariant visible sector
will feel no SUSY breaking at tree-level from supergravity VEVs, in particular ${\cal F}_{C_0}$, that is no breaking of sequestering. But we must consider higher (super-)derivative couplings of $C_0$ to the visible sector, that might arise in the 4D effective theory upon careful matching to the 5D theory. Rather than doing such a matching, we again place crude bounds: extra superderivatives are suppressed by powers of at least the cutoff of the 4D EFT, $m_{radion}$, so that this source of visible soft mass-squareds is also suppressed by the dimensionless $\Lambda^2/m_{radion}^2$, compared to the going rate. (If the superderivatives were not required we would have gotten soft terms of order $|{\cal F}_{C_0}|^2 \propto \Lambda^4$.) Again, we will find this suppression is enough to neglect this source of visible SUSY breaking compared to AMSB.

This observation that $\Lambda^2/m_{radion}^2$ suppresses non-sequestered contributions is the main value added by the effective theory organization of corrections.

\section{Bulk Radiative Corrections}
\label{BRC}

The spectrum of pure AMSB applied to MSSM has a serious problem, namely the negative mass squared of the sleptons. However, this can be compensated by other positive flavor-conserving contributions generated by bulk gravity loops. However, there are other types of bulk loops that can violate flavor symmetry such as the hypermultiplet loops in our model. We must ensure that the gravity loops can be of the right size and sign to correct slepton masses while hypermultiplet loops are far subdominant, so as not to reintroduce supersymmetric flavor problem.

The flavor-blind result of the gravity loops to visible sector scalar masses was calculated in Ref.~\cite{RattazziSS:2003,Buchbinder:2003qu,GregoireRS:2005},
\begin{equation}
\label{b2bgravity}
   (m^2_{vis})^{gravity} \sim  \frac{1}{16\pi^2} \frac{\Lambda^4}{M^2_4} \frac{1}{(\pi r)^3}.
\end{equation}
Note that $\Lambda^4$ represents the supersymmetry breaking vacuum energy as was discusssed in Section \ref{subHSSB}. The result is UV-finite because of the finite separation between the visible and hidden sectors. We can view this result as the quadratically divergent result one would obtain in a purely 4D theory, but with the UV-cutoff replaced by a ``KK regularization" $1/\pi r$. This contribution is positive in a class of a supersymmetry breaking models with a dominant hidden $D$-term~\cite{GregoireRS:2005}.

In an analogous fashion, hypermultiplet loops also mediate between the hidden and visible sectors via the couplings in the action,
\begin{equation}
\label{hypercoupl}
   \int d^4 x\int d\varphi \int d^4 \theta\, \left ( \Lambda^4 \theta^2\bar\theta^2 H^*H\delta(\varphi) + c_{ij}\, Q_iQ^*_j HH^*\delta(\varphi-\pi) \right ).
\end{equation}
Modifying the highly warped estimates of Ref.~\cite{MaruO:2004}, we will simply estimate the hypermultiplet loop effect as the quadratically divergent contribution from the 4D effective low energy theory, with the resulting UV-cutoff replaced by $1/\pi r$. In the 4D effective theory $H$ in Eq. (\ref{hypercoupl}) is replaced by its zero mode given by Eq. (\ref{rigidsol}) (with $T=\pi r$). Thus the hypermultiplet induced visible scalar masses, in general flavor-violating, are estimated to be 
\begin{equation}
\label{b2bhyper}
\begin{split}
  m^{2}_{vis} \sim \frac{1}{16 \pi^2} \frac{4\, e^{-2m_0 \pi r}}{(1- e^{-2m_0 \pi r})^2}\, \frac{m^2_0}{\pi r} \frac{\Lambda^4}{M^2_4}\, .
\end{split}
\end{equation}
Note that $H_0$ is not canonically normalized, but we have taken its wave function into account. We expect this to be a conservative estimate.

\section{Numerical Estimates}
The bulk radiative corrections from the supergravity fields given by Eq. (\ref{b2bgravity}) can compensate the negative slepton mass-squared if they are comparable in magnitude. For example, it is sufficient if roughly
\begin{equation}
\label{b2bgravitycons}
    (m^2_{vis})^{\ gravity} \gtrsim 3 \times 10^{-5}\, \frac{\Lambda^4}{M^2_4}\, .
\end{equation}
To adequately suppress flavor-violating contributions it is sufficient that roughly~\cite{Luty:2001jh}
\begin{equation}
\label{b2bhypercons}
    (m^{2}_{vis})^{\ flavor-violating}  \lesssim 3 \times 10^{-7}\, \frac{\Lambda^4}{M^2_4}\, .
\end{equation}
Saturating Eq. (\ref{b2bgravitycons}) using Eq. (\ref{b2bgravity}) roughly determines the size of the extra-dimension to be
\begin{equation}
\label{fixsize}
\pi r \sim 6\, .
\end{equation}
Given the value of $\pi r$ and using $M^2_4 =\pi r$, the 5D hypermultiplet mass, $m_0$, can be adjusted in Eq. (\ref{b2bhyper}) to satisfy the constraint given by Eq. (\ref{b2bhypercons}).

As was discussed in Section \ref{4ETS}, sequestering is guaranteed in a simple way by requiring the radion to be heavy, $\sim 10^{15}$ GeV in order to satisfy Eq. (\ref{b2bhypercons}) with our sequestering factor $\sim \Lambda^2/m^2_{radion}$.
Given Eq. (\ref{fixsize}) and 4D Planck mass $M_4 = \pi r \sim 10^{18}$ GeV, the radion mass in a 5D Planck unit is given by
\begin{equation}
\label{mradplanck}
     m_{radion} \sim 2.4 \times 10^{-3}\, .
\end{equation}
The expression of the radion mass is given by Eq. (\ref{radionmass}). For fixed $m_0$ and $\pi r$, $J_0$ can be adjusted to get the value given by Eq. (\ref{mradplanck}).

A possible viable choice of parameters is:
\begin{equation}
   m_0 = 7.5 \times 10^{-1}\, ,\quad J_0 = 1.9 \times 10^{-3}\, , \quad J_\pi = 1.7 \times 10^{-1}\, .
\end{equation}
This choice gives rise to $e^{-2m_0\pi r} \sim 1.2 \times 10^{-4}$, which is the key to ensuring the suppression of hypermultiplet loop contributions to Eq. (\ref{b2bhypercons}) relative to Eq. (\ref{b2bgravitycons}), that is the suppression of Eq. (\ref{b2bhyper}) relative to Eq. (\ref{b2bgravity}). Note that $e^\sigma \sim 1$ for the above choice of parameters (Recall that $\sigma$ is determined by Eq. (\ref{warpfactor}) with the boundary condition, $\sigma(\varphi=0)=0$).

\section*{Acknowledgments}
We would like to thank Jonathan Bagger, Dmitry V. Belyaev, Andrew E. Blechman, David E. Kaplan, Taichiro Kugo, Markus A. Luty, Keisuke Ohashi for discussions and advice. M.S. especially thanks Gero von Gersdorff for many discussions and technical assistance. The research of R.S. and M.S. was supported by the National Science Foundation grant NSF-PHY-0401513 and by the Johns Hopkins Theoretical Interdisciplinary Physics and Astrophysics Center.

\appendix

\section{Convention}
\label{ac}
Throughout the paper, we will use $A,B,...= \dot 0, \dot 1, \dot 2, \dot 3, \dot 5$ for the flat 5D spacetime indices and $M,N,... = 0,1,2,3,5$ for the curved 5D indices. Similarly we will use $a,b,...= \dot 0, ..., \dot 3$ for the flat 4D spacetime indices and $\mu,\nu,... = 0,...,3$ for the curved 4D indices. We mainly follow the convention of Ref.~\cite{KugoMatter:2002,KugoOrbifolding:2002}.
The convention of the metric is
\begin{equation}
   \eta_{AB} = {\rm diag}(1,\, -1,\, -1,\, -1,\, -1 ).
\end{equation}
The gamma matrices are
\begin{equation}
\gamma ^\mu = \begin{pmatrix} 0 & \sigma^\mu \\ \bar \sigma^\mu & 0 \end{pmatrix}, \quad \quad
\gamma ^{\dot 5}=\begin{pmatrix} -i & 0 \\ 0 & i \end{pmatrix}\, ,
\end{equation}
where $\sigma^\mu =(-1, \vec\sigma)$ and $\sigma ^{\bar\mu}=(-1, -
\vec\sigma)$. The gamma 'five' matrix, $\Gamma_5$, is defined as $\Gamma_5 \equiv  i \gamma ^{\dot 5}$ and the projection operaters are defined as
\begin{equation}
   {\cal P}_R = \frac{1}{2}(1 + \Gamma _5 ), \quad \quad 
   {\cal P}_L = \frac{1}{2}(1 - \Gamma _5 ).
\end{equation}
The antisymmetrized gamma matrices are
\begin{equation}
   \gamma^{A_1A_2\cdots A_n} = \gamma^{[A_1}\gamma^{A_2}\cdots \gamma^{A_n]} \equiv
   \frac{1}{n!}\sum_{\rm perms}(-1)^p\gamma^{A_1}\gamma^{A_2}\cdots\gamma^{A_n}.
\end{equation}
The reality constraint of the scalar of the hypermultiplet ( Similarly for the straight $F$ terms) is
\begin{equation}
  A^i_\alpha \equiv - A^{\alpha *}_i = \epsilon^{ij} A^\beta_j \rho_{\beta\alpha}.
\end{equation}
The convention of the $\epsilon$ tenser is $\epsilon_{12}=\epsilon ^{12} =1$. The summation convention is from the northwest to southeast, i.e. $\bar\psi\chi \equiv \bar\psi^i\chi_i$. The indices are raised or lowered according to the rules, $A^i = \epsilon^{ij}A_j$ and $A_i = A^j\epsilon_{ji}$.
We parameterize the scalar fields and F-temrs as
\begin{equation}
A^{\alpha =3,4} _i =
\begin{pmatrix} A^{\alpha =3}_{i=1}= H^* & A^{\alpha =3}_{i=2}=\bar H  \\ A^{\alpha =4}_{i=1}= -\bar H^*  & A^{\alpha =4}_{i=2}=H  \end{pmatrix}, \quad \quad 
A^{\alpha=1,2} _i =
\begin{pmatrix} A^{\alpha =1}_{i=1}= C^* & A^{\alpha =1}_{i=2}= \bar C \\ A^{\alpha =2}_{i=1}= -\bar C^* & A^{\alpha =2}_{i=2}= C \end{pmatrix},
\end{equation}
\begin{equation}
F^{\alpha=3,4} _i =
\begin{pmatrix} F_{\bar H} & - F_H^* \\ F_H & F_{\bar H}^* \end{pmatrix}, \quad \quad 
F^{\alpha=1,2} _i =
\begin{pmatrix} F_{\bar C} & - F_C^* \\ F_C &  F_{\bar C}^* \end{pmatrix}.
\end{equation}
The $C$ denotes the 5D compensator and the gauge fixing of the $SU(2)$ $\bf U$ symmetry in a bulk theory is done by choosing $A^{\alpha=1,2}_i \propto \delta^\alpha_i$. 
The reality constraint of the Symplectic-Majorana spinors is given by
\begin{equation}
\label{arcf}
   {\bar\zeta}^\alpha \equiv (\zeta _\alpha)^\dagger \gamma ^0 = {\zeta^\alpha}^T {\bold C},
\end{equation}
where ${\bold C}$ is the 5D charge conjugation matrix. We parameterize the Symplectic-Majorana spinors, satisfying Eq. (\ref{arcf}), as
\begin{equation}
\zeta^1 = \begin{pmatrix} \zeta_+ \\ - \bar \zeta_- \end{pmatrix}, \quad 
\zeta^2 = \begin{pmatrix} \zeta_- \\ \bar \zeta_+ \end{pmatrix}, \quad
{\bar \zeta}^1 = \begin{pmatrix} \zeta _+ & \bar \zeta _- \end{pmatrix}, \quad
{\bar \zeta}^2 = \begin{pmatrix} \zeta _- & - \bar \zeta _+ \end{pmatrix}\, .
\end{equation}
4D Majorana fermions can be obtained out of 
\begin{equation}
   \zeta = \zeta^1_R + \zeta^2_L = \begin{pmatrix} \zeta _+ \\ \bar \zeta _+ \end{pmatrix}, \quad \quad 
   \zeta' = i( \zeta^1_L + \zeta^2_R ) = i \begin{pmatrix} \zeta _- \\ -\bar \zeta _- \end{pmatrix}.
\end{equation}
The $\zeta$ and $i \Gamma^5 \zeta '$ become 4D Majorana fermions.
The matrix $d^{\ \alpha}_\beta$ appearing in the kinetic terms, Eqs. (\ref{Kugobulkpot}) ,(\ref{Kugodef}) in the Appendix \ref{BP4SSN}, of the 5D hypermultiplet is given by
\begin{equation}
\label{dmat}
    d^{\ \alpha}_\beta = \begin{pmatrix} {\bold 1}_2 &0 \\ 0 & -{\bold 1}_2 \end{pmatrix}.
\end{equation}
One notes that ${\bold 1}_2$ corresponds to the physical hypermultipelt while  $-{\bold 1}_2$ to unphysical compensator. Another type of matrix, $\eta_{\alpha\beta}$, appears in the mass terms of the 5D hypermultiplet ( See 
Eq. (\ref{Kugodef}) in Appendix \ref{BP4SSN} ). The $\eta$ matrix must be symmetric, or $\eta_{\alpha\beta} = \eta_{\beta\alpha}$. The diagonal components of $\eta$ matrix must vanish not to break $U(1)_R$ symmetry.
The isovector, e.g. $\vec Y = (Y^1, Y^2, Y^3)$ is related to $SU(2)$ tensor, $Y^{ij}$, by the relation,
\begin{equation}
   Y^i_{\, j} = Y^{ik}\epsilon_{kj} = i \vec Y \cdot \vec \sigma^i_{\, j}\, ,
\end{equation}
and satisfies hermiticity $Y^{ij}=(Y_{ij})^*$.

\section{Bulk Potential for 4D Scalars in Superspace Notation}
\label{BP4SSN}
We collect only terms contributing to the scalar potential for 4D scalars from Eq. (4.7) in Ref.~\cite{KugoMatter:2002} (Note that $m \rightarrow  2i\,m$ has been made.) 
\begin{equation}
\label{Kugobulkpot}
\begin{split}
  -V_{bulk} =\int d\varphi\, e\, &\left \{ -\frac{1}{2}{\cal N}{\cal R}_5 - {\cal N}V_5^{ij}V^5_{ij} -\frac{1}{2}{\cal N}_{00}Y^{0ij}Y^0_{ij} \right . \\
   &- 2 V^5_{ij}A^{\beta i}d^{\, \alpha} _\beta \nabla _5 A_\alpha^j + 2Y^0_{ij} A^i_\alpha(gt_0)^{\bar \alpha \beta} A^j_\beta \\
   &+ \nabla ^5 A^{\bar\alpha}_i \nabla _5 A^i_\alpha - b^5\partial _5 {\cal N} - \frac{1}{4}({\cal N}_{00} - \frac{{\cal N}_0{\cal N}_0}{{\cal N}})\partial _5 M \partial^5 M \\
   &\left . +(1 - \frac{A_5 A^5}{M^2})\tilde F^{\bar \alpha}_i \tilde F_\alpha ^i + A^{\bar\alpha}_i (gMt_0)^2 A^i_\alpha + {\cal C}'({\cal A}^2 + 2{\cal N}) \right \} \,  .
\end{split}
\end{equation}
where
\begin{equation}
\label{Kugodef}
\begin{split}
    {\cal N} &= M^3, \quad b^5 = M^{-1}\partial ^5 M, \quad {\cal A}^2 = A^{\bar\alpha}_i A^i_\alpha \equiv  A^{\beta}_i d^{\alpha}_{\beta} A^i_\alpha \, ,\\
    (gM t_0)^\alpha_\beta A^\beta_i &= -M(gt_0)^{\alpha\beta}A_{\beta i} = im M\eta^{\alpha\beta}A_{\beta i}\, , \\
   (gMt_0)^2 A^i_\alpha &= (gMt_0)_{\alpha\gamma} (gMt_0)^\gamma_\beta A^{\beta i }\, , \\
   \tilde F^\alpha_i &= F^\alpha_i + im M(d^{-1})^\alpha_\gamma \eta^{\gamma\beta}A_{\beta i}\, , \\
   \nabla_5 A^i_\alpha &= \partial _5 A^i_\alpha - A_5\,(gt_0)_{\alpha\beta} A^{\beta i}\, .
\end{split}
\end{equation}
Note that instead of using $t^{ij}$ to refer to auxiliary field that belongs to the central charge vector multiplet, we used $Y^{0\, ij}$ to refer to same field (See Ref.~\cite{FujitaKO:2001} for similar discussion). Throughout the paper, we have used $Y$ (instead of $Y^0$) since our set up has only one central charge vector multiplet. The relation of ${\cal C}'$ to unprimed ${\cal C}$ as well as the full details of the bulk action is given in the same Ref.~\cite{KugoMatter:2002}.

Since we are interested in working in an orbifolded superconformal gravity theory on an $S^1/Z_2$ and the orbifolding breaks 5D superconformal gauge symmetry down to 4D superconformal gague symmetry, we will re-write the above bulk scalar potential given by Eq. (\ref{Kugobulkpot}) in terms of the fields transforming like representations of 4D $N=1$ supersymmetry. We follow the convention given in Appendix \ref{ac}.
\begin{equation}
\label{KugobulkpotZ2}
\begin{split}
  -{V}_{bulk} = \int d\varphi\, e\, &\left \{ - \frac{1}{2}M^3\, {\cal R}_5 - 6 M\, \vec Y\cdot \vec Y - \frac{3}{2}M\, \partial _5 M \partial^5 M \right . \\ 
  & + 4 m( H\bar H Y + H^*\bar H^* Y^* + Y^3(|H|^2 - |\bar H|^2) ) \\
  &+ 2\, D_5 H^* D^5H + 2\, D_5 \bar H^* D^5\bar H  - 2\, D_5 C^* D^5 C  - 2\, D_5 \bar C^* D^5 \bar C \\
  & - 2m^2 A_5 A^5 (|H|^2+|\bar H|^2) \\
  &- 2 (1 - \frac{A_5 A^5}{M^2}) \{ |F_C|^2 + |F_{\bar C}|^2 - |F_H|^2 - |F_{\bar H}|^2 \\
  & + i\, m M( F_{\bar H} H + F_H \bar H - F_H^* \bar H^* - F_{\bar H}^*H^* ) \} \\
  &\left . + {\cal C}'({\cal A}^2 + 2M^3) \right \} \, .
\end{split} 
\end{equation}
The covariant derivatives are given by
\begin{equation}
\begin{split}
  D_5 H &= ( \partial _5  - i V^3_5 - i\, m A_5 )H - iV^*_5\bar H^*\, ,
\\
  D_5 \bar H &= (\partial _5 - i V^3_5 + i\, m A_5 )\bar H + i V^*_5H^* \, ,
\\
  D_5 C &= ( \partial _5  - i V^3_5)C - iV^*_5\bar C^*\, ,
\\
  D_5 \bar C &= (\partial _5 - i V^3_5 )\bar C + i V^*_5 C^* \, .
\end{split}
\end{equation}
The $\vec Y\cdot \vec Y$ denotes the inner product of isovectors $\vec Y = (Y^1,Y^2,Y^3)$ (same for $\vec V_M$).
It is convenient to complexify parity even components $Y^{1,2}, V^{1,2}_5$ as $Y \equiv Y^1+iY^2$, $V_5 \equiv V_5^1+i V^2_5$. The matrix elements, $\eta_{34}=\eta_{43}=\eta$, appearing in Eq. (\ref{Kugodef}) was absorbed into $m$, i.e. $m\, \eta \rightarrow m$. Using the choice for the matrix $d^{\ \alpha}_\beta$ given in Eq. (\ref{dmat}) the ${\cal A}^2$ is given by 
\begin{equation}
 {\cal A}^2 \equiv A^\beta_i d_\beta^\alpha A^i_\alpha = 2(-|C|^2 -|\bar C|^2 + |H|^2 + |\bar H|^2)\, .
\end{equation}
One notes the equation of motion of ${\cal C}'$ gives rise to one constraint, ${\cal A}^2 = -2M^3$.

On the 5D warped spacetime metric,
\begin{equation}
     ds^2 = e^{2\sigma}\eta_{\mu\nu}dx^\mu dx^\nu - r^2d\varphi^2\, ,
\end{equation}
the ``$e$" and the 5D curvature term ${\cal R}_5$ in Eq. (\ref{KugobulkpotZ2}) are given by ( See Refs.~\cite{CorreiaST:2005,AbeS:2004sd} for related discussions)
\begin{equation}
 e = e^{4\sigma}\, r\, , \quad {\cal R}_5 = 4\,(2\, \partial_{\dot 5}^2 \sigma + 5\, ({\partial_{\dot 5} \sigma})^2)\, .
\end{equation}
The warp factor-dependence in a basis with the unit warp factor is easily determined by observing that under the Weyl transformation,
\begin{equation}
    e^a_{\ \mu} \rightarrow e^{-\sigma} e^a_{\ \mu}\, ,
\end{equation}
the fields transform like~\cite{CorreiaST:2005}
\begin{equation}
\begin{split}
   H &\rightarrow e^{3/2\, \sigma} (H, e^{\sigma/2}\zeta_H, e^\sigma {\cal F}_H) \equiv e^{3/2\, \sigma } H\, , \\
   {\bf V} &\rightarrow e^\sigma (A_\mu, 2 e^{\sigma/2} \Omega_+, e^\sigma D_{{\bf V}})\equiv e^\sigma  {{\bf V}}\, , \quad D_{{\bf V}} = 2Y^3 - \partial _{\dot 5} \sigma - \hat {\cal D}_{\dot 5}M\, , \\
   -2i\, {\bf\Sigma} &\rightarrow 
   (e^{\dot 5}_5 M - i A_5, e^{\sigma/2} (-4i\, \psi_{5-} M + 4e^{\dot 5}_5 \Gamma_5 \Omega_- ) , e^\sigma (-2i(V^1_5 + i V^2_5)M \\ 
  &\quad - 2e^{\dot 5}_5 (Y^1 + i Y^2) + 2i \psi_{5-}(1+ \Gamma _5)\Omega _-) ) \equiv -2i\,  {\bf \Sigma}\, .
\end{split}
\end{equation}
The effective potential from the bulk action given by Eq. (\ref{KugobulkpotZ2}) can be economically written in compact flat (global) 4D superspace notation,
\begin{equation}
\label{superspacepot}
\begin{split}
 - V_{bulk} = - \int d\varphi &\left \{ \int d^4\theta\,e^{2\sigma}\,\frac{1}{2}\, (T+T^*)\, \left (\,2CC^* + 2\bar C\bar C^* - H^* e^{\frac{2}{\sqrt{3}}m\, \tilde V} H - \bar H e^{-\frac{2}{\sqrt{3}} m\, \tilde V}\bar H^* + \frac{1}{2} \tilde {\cal V}^3 \right ) \right . \\
   & + \int d^2 \theta\,e^{3\sigma}\,\left ( H(\frac{1}{2}\, \stackrel{\leftrightarrow}{\partial _5} +\, 2i\, m\Sigma)\bar H - C \stackrel{\leftrightarrow}{\partial _5} \bar C + \frac{i}{2}\, \Sigma\, \tilde{\cal W}_\alpha \tilde{\cal W}^\alpha + {\rm  h.c.} \right )\\
   &+ e^{4\sigma}r\, \left \{ -6\,M |Y|^2 +  V^3_5 V^{3\, 5} {\cal A}^2 \right . \\
   &\hspace{1.5cm}\left . + (\, 2\,m\,H\bar H Y + i V_5^3 (\, H^*\nabla ^5 H + \bar H^*\nabla^5\bar H  - 2 C^* \partial ^5 C - 2 \bar C^* \partial ^5 \bar C \, ) + {\rm h.c.} )\right \}\\
   & \left .+ e^{4\sigma}r\,{\cal C}'({\cal A}^2 + 2M^3) \right \} \, ,
\end{split}
\end{equation}
in terms of the ``fake" flat superspace multiplets,
\begin{equation}
\label{appMdef}
\begin{split}
  \tilde V &= \frac{1}{2}\theta^2\bar\theta^2(\sqrt{3}\, e^{2\sigma} D)\, , \quad \tilde{\cal W}_\alpha = \theta\, e^{\sigma/2}\,\sqrt{3}\, D\, , \\
  \quad \tilde {\cal V} &= M - \theta^2\bar\theta^2  \partial_{\dot 5} ( e^{2\sigma}\,D)\, ,\quad e^\sigma D \equiv e^\sigma\, 2Y^3-\partial_{\dot 5} (e^\sigma M)\, ,\\
   T &= \frac{-2i}{M}\Sigma = T + \theta^2 e^\sigma {\cal F}_T\, , \quad T \equiv r -\frac{i}{M}A_5\, ,  \quad {\cal F}_T \equiv -2i\,V_5\, , \\
   C &= C + \theta^2 e^\sigma {\cal F}_C \, ,\quad  \bar C = \bar C + \theta^2 e^\sigma {\cal F}_{\bar C}\, , \quad 
   H = H + \theta^2 e^\sigma {\cal F}_H\, , \quad \bar H = \bar H + \theta^2 e^\sigma {\cal F}_{\bar H}\, , \\
  {\cal A}^2 &= 2(-|C|^2 - |\bar C|^2 + 1/2\, |H|^2 + 1/2\,|\bar H|^2)\, .
\end{split}
\end{equation}
The ${\cal F}$ terms are defined in terms of 5D $F$ terms and other 5D fields as
\begin{equation}
\label{appFdef}
\begin{split}
   {\cal F}_H &= \frac{1}{r} \, \{ - \partial_5 \bar H^* - \frac{3}{2}\partial_5 \sigma\, \bar H^* + i ( r + \frac{i}{M}A_5)F_H + i(V^1_5+iV^2_5)H \}\, , \\
   {\cal F}_{\bar H} & = \frac{1}{r} \, \{ \partial_5 H^* + \frac{3}{2}\partial_5 \sigma\, H^* + i ( r + \frac{i}{M} A_5)F_{\bar H} + i(V^1_5+iV^2_5)\bar H  \}\, , \\
   {\cal F}_C &= \frac{1}{r}  \, \{-\partial _5 \bar C^* -\frac{3}{2}\partial _5\, \sigma \bar C^* + i ( r + \frac{i}{M}A_5)F_C + i(V^1_5+iV^2_5)C \}\, , \\
   {\cal F}_{\bar C} & = \frac{1}{r}  \, \{ \partial_5 C^* + \frac{3}{2}\partial_5 \sigma\, C^* + i ( r + \frac{i}{M}A_5)F_{\bar C} + i(V^1_5+iV^2_5)\bar C  \}\, .
\end{split}
\end{equation}
The covariant derivatives are defined as
\begin{equation}
   \nabla_5 H = \partial _5 H - im A_5 H\, , \quad \nabla_5 \bar H = \partial _5 \bar H + i m A_5 \bar H\, .
\end{equation}
Note that the $(H, \bar H) \rightarrow \frac{1}{\sqrt{2}}\, (H, \bar H)$ was made (same for 5D $F$ terms of $H$ and $\bar H$ ), compared to original component bulk scalar potential of Eqs. (\ref{Kugobulkpot}) and (\ref{KugobulkpotZ2}), so that physical hypermultiplet has a canonical normalization. 

By performing superspace integration of the bulk potential given by Eq. (\ref{superspacepot}) with the expressions given by Eq. (\ref{appMdef}) and (\ref{appFdef}), one can easily reproduce all the component terms in Eq. (\ref{KugobulkpotZ2}). Note that the bulk potential given by Eq. (\ref{superspacepot}) has the explicit dependence of dilaton $M$ and $\bar C$ which means that we have not yet gauge-fixed the scale invariance and $SU(2)$ ${\bf U}$ symmetry.

\end{document}